# A Fully Automated and Explainable Algorithm for the Prediction of Malignant Transformation in Oral Epithelial Dysplasia

Adam J Shephard[1*], Raja Muhammad Saad Bashir[1], Hanya Mahmood[2], Mostafa Jahanifar[1], Fayyaz Minhas[1], Shan E Ahmed Raza[1], Kris D McCombe[3], Stephanie G Craig[3], Jacqueline James[3], Jill Brooks[4], Paul Nankivell[4], Hisham Mehanna[4], Syed Ali Khurram[2‡], Nasir M Rajpoot[1‡]

[1] *Tissue Image Analytics Centre, Department of Computer Science, University of Warwick, Coventry, UK*

[2] *School of Clinical Dentistry, University of Sheffield, Sheffield, UK*

[3] *Precision Medicine Centre, Patrick G. Johnston Centre for Cancer Research, Queen's University Belfast, UK*

[4] *Institute of Head and Neck Studies and Education, Institute of Cancer and Genomic Sciences, University of Birmingham, UK*

\* adam.shephard@warwick.ac.uk

[‡] Joint co-senior authorship

## Summary

**Background:** Oral epithelial dysplasia (OED) is a premalignant histopathological diagnosis given to lesions of the oral cavity. Its grading suffers from significant inter-/intra- observer variability, and does not reliably predict malignancy progression, potentially leading to suboptimal treatment decisions. To address this, we developed a deep learning pipeline for objectively predicting OED progression to malignancy in Haematoxylin and Eosin (H&E) stained whole slide images (WSIs).

**Methods:** We have developed a novel artificial intelligence algorithm that can assign an Oral Malignant Transformation (OMT) risk score based on histological patterns in the H&E WSIs to quantify the risk of OED progression. The algorithm is based on the detection and segmentation of nuclei within (and around) the epithelium using an in-house segmentation model. We then employed a shallow neural network fed with interpretable morphological/spatial features, emulating histological markers. We conducted internal cross-validation on our development cohort (Sheffield; $n = 193$ cases) followed by independent validation on two external cohorts (Birmingham and Belfast; $n = 92$ cases).

**Findings:** The proposed *OMTscore* yields an AUROC = 0.74 in predicting whether an OED progresses to malignancy or not for internal and external cohorts. Survival analyses showed the prognostic value of our *OMTscore* for predicting malignancy transformation, when compared to the manually-assigned WHO and binary grades. Nuclear analyses of the correctly predicted cases elucidated the presence of peri-epithelial and epithelium-infiltrating lymphocytes in the most predictive patches of cases that transformed ($p < 0.0001$).

**Interpretation:** This is the first study to propose a completely automated algorithm for predicting OED transformation based on interpretable nuclear features, whilst being validated on external datasets. The algorithm shows better-than-human-level performance for prediction of OED malignant transformation and offers a promising solution to the challenges of grading OED in routine clinical practice.

**Funding:** Cancer Research UK Early Detection Project Grant (C63489/A29674).

## Introduction

Head and neck cancer is among the top ten most common cancers worldwide[1], with approximately 150,000 new cases diagnosed annually in Europe[2]. These cancers are often detected at an advanced stage (~60%), resulting in poor prognosis and a five-year survival rate of only 40%[2]. With early diagnosis followed by timely treatment, survival increases to 80-90%[2]. Therefore, early detection plays a crucial role in improving patient outcomes.

Oral squamous cell carcinoma (OSCC) is the most common type of head and neck cancer[1], that may arise from an oral potentially malignant disorder (OPMD) such as leukoplakia or erythroplakia[3]. These disorders are often associated with lifestyle habits such as tobacco smoking, betel quid chewing and excessive alcohol consumption, although genetic factors may also play a role[4–6]. Following a biopsy and microscopic examination, these lesions may be given a histopathological diagnosis of oral epithelial dysplasia (OED), which carries a higher risk of progressing to OSCC[4,7]. Histological atypia in OED typically manifests in the basal layer and progresses upwards through the epithelial layers. Cytological changes often include changes to the shape, size and colour of nuclei/cells, the presence of atypical mitotic figures, and increased cellularity[3]. Architectural changes typically include irregular epithelial stratification, loss of basal cell polarity, drop-shaped rete pegs, and loss of epithelial cohesion[3].

There are different grading systems to classify OED and inform treatment decisions. The 2017 World Health Organisation (WHO) grading is a three-tier system for grading cases as mild, moderate and severe, taking into account over 15 different features. This system splits the epithelium into thirds, suggesting that architectural/cytological changes confined to the lower third may be classed as mild, in the middle moderate, and those progressing towards the upper third as severe[8]. However, this system oversimplifies a complex disease process, lacks standardisation, and introduces ambiguity and subjectivity, which could result in an inaccurate diagnosis with potentially detrimental implications for outcomes. An alternate binary grading system, categorising lesions as low- or high-risk, based on the number of cytological and architectural features, aimed to improve grade reproducibility[8,9]. However, studies have shown significant variability in grading using both systems[3], highlighting the need for a more objective and reproducible methods that can better predict malignant transformation in OED.

The availability of graphical processing units (GPU) and the rise of convolutional neural networks (CNNs) and deep learning have revolutionised computer vision, including medical imaging[10]. Computational pathology is an active area of research that leverages machine learning and deep learning algorithms for the analysis of histological patterns in multi-gigapixel whole-slide images (WSIs) to tackle pathology related tasks[11,12]. Deep learning models have become commonplace in laboratories worldwide, being used for tasks such as segmentation, detection, and classification[13–16]. Numerous deep learning algorithms have been applied to tasks such as tissue and nuclei segmentation in WSIs[17–23], as well as making slide-level predictions for histopathological diagnoses[24–27]. Multiple studies have proposed generating slide-level predictions by aggregating patch-level predictions or features using pooling or attention-based mechanisms[24,28–31]. Efforts are underway to consolidate the diverse deep learning methods employed in computational pathology, exemplified by initiatives like the *TIAToolbox*[32].

Several studies have explored the use of AI in grading and prognostication of OED lesions. Bashir *et al.* (2020)[23] used the mean widths of epithelial layers as a proxy for epithelial stratification, within Random Forests to predict OED grade. Shephard *et al.* (2022)[26] achieved varying success in predicting OED recurrence/transformation using nuclear shape/size features in H&E images.

Mahmood *et al.* (2022)[33] employed pathologist-derived features in Cox proportional hazards regression models to predict recurrence and transformation, identifying prognostic features such as bulbous rete pegs, hyperchromatism, and nuclear pleomorphism. Although manual feature extraction was required, the study demonstrated the link between OED features and clinical outcome. In contrast, Bashir *et al.* (2023)[34] used weakly supervised multiple instance learning and identified peri-epithelial lymphocytes (PELs) as a prognostic feature for transformation at the WSI-level. However, this method required manually refined epithelial masks, and its success was not validated on external datasets. These studies demonstrate the potential of AI in improving OED diagnosis and prognosis but also emphasise the need for further development and validation of fully automated methods.

In this study, we present an end-to-end fully automated and explainable pipeline for predicting OED transformation. We utilise an in-house multi-task model[18] to generate nuclear and intra-epithelial layer segmentations and extract morphological/spatial features. These features are then fed into a multi-layer perceptron (MLP) to predict slide-level malignant transformation of OED. Our contributions to the scientific community include:

1. Introduction of our pipeline's automatically generated *OMTscore*, to improve diagnostic OED grading. Validation of the *OMTscore* was conducted on independent cohorts from Birmingham and Belfast, UK.
2. Presentation of HoVer-Net+, a state-of-the-art model capable of simultaneous segmentation and classification of nuclear instances and intra-epithelial layers. We have released the model code and weights as part of the TIAToolbox[32].
3. Demonstrated the superiority of our *OMTscore* over conventional histological grading in predicting malignancy transformation. Our code for model inference is publicly accessible at: [adamshephard/OMTscoring_inference (github.com)](github.com).

## Materials and Methods

**Study Data**

*Internal Training/Validation*

The study cohort used for training our models consisted of subjects collected retrospectively between 2008 and 2016 from the Oral and Maxillofacial Pathology archive at the School of Clinical Dentistry, University of Sheffield, UK. Ethical approval was obtained from the NHS Health Research Authority West Midlands (18/WM/0335). Sections were newly cut (4 µm thickness) and H&E stained from formalin fixed paraffin embedded blocks.

The dataset comprised 279 slides with a histological diagnosis of OED, scanned using either a Hamamatsu NanoZoomer 360 (Hamamatsu Photonics, Japan) or an Aperio CS2 (Leica Biosystems, Germany) digital slide scanner at 40× objective power (0.2258 mpp and 0.2520 mpp, respectively) to obtain digital WSIs. Clinical information was available for 270 slides, including patient age, sex, intraoral site, OED grade (binary and WHO 2017), and transformation status. Transformation was defined as the progression of a dysplastic lesion to OSCC at the same clinical site within the follow-up period. Multiple certified/consultant pathologists independently evaluated the cases when initially reported using the WHO grading system (PMS, PMF, DJB), to ensure diagnostic consistency. Blind re-evaluation was performed by an Oral & Maxillofacial Pathologist (SAK) and an Oral Surgeon specialising in OED analysis (HM), to confirm the WHO (2017) grade and assign binary grades. The cohort included 193 unique OED patients with 42 patients (57 slides) exhibiting malignant transformation. Slides from the same patients were consistently assigned to the same fold during training/internal cross-validation. A summary of the cohort is provided in Table 1.

For training our segmentation models, one expert pathologist (SAK) exhaustively manually delineated the intra-epithelial layers (basal, epithelial, and superior keratin layers) in 59 OED cases, in addition to nine controls (collected with the Aperio CS2 scanner as per the above protocols), using our in-house WASABI software (a customised version of HistomicsTK[35]). We then generated tissue masks for each of the segmented WSIs via Otsu thresholding and the removal of small objects and holes in the segmentation mask. A layer mask was then generated for each WSI by combining the layer segmentations with the tissue mask.

Nuclear instance masks were generated for 30 regions of interest (one ROI per case) where a pathologist (SAK) annotated each nucleus as either epithelial or 'other'. The point annotations were used within the NuClick framework to generate nuclear boundaries[19,36]. NuClick has been found to be superior to fully automated approaches for generating nuclear instance segmentations, particularly in the cases of touching/overlapping nuclei[19]. To ensure that all nuclear segmentations were of a high quality, the masks were then manually refined when found to be visibly incorrect. A total of 71,757 labelled nuclei segmentations were obtained across the 30 ROIs, which were used to train our segmentation models.

*Independent External Validation*

For external validation, OED cases from two independent centres, Birmingham and Belfast, were recruited. A total of 47 OED patients' data were collected from Belfast and 69 OED cases were collected from Birmingham. The Birmingham and Belfast slides were scanned at 40× objective power using a Pannoramic 250 (3DHISTECH Ltd., Hungary; 0.1394 mpp) and an Aperio AT2 (Leica Biosystems, Germany; 0.2529 mpp) scanner, respectively. The cases were assigned mild, moderate or severe WHO grades (WHO 2017), and additionally had time to transformation data. The combined Birmingham-Belfast external validation cohort consisted of 116 unique OED cases, with 42 cases transitioning to malignancy. A summary of these cohorts is provided in Table 1.

**Table 1**

*Demographic, characteristic and outcome data for the OED cases from Sheffield, Birmingham and Belfast cohorts.*

|  | Sheffield | Belfast | Birmingham |
| --- | --- | --- | --- |
| *n* | 270 | 47 | 69 |
| Median Age (IQR) | 64 (55 – 74) | 61 (51 – 69) | 61 (51 – 70) |
| Sex, *n* (%) | | | |
|     Female | 124 (46) | 23 (49) | 30 (44)* |
|     Male | 145 (54) | 24 (51) | 38 (56) |
| Site, (%) | | | |
|     Buccal Mucosa | 35 (13) | 0 (0) | 12 (17) |
|     Tongue | 118 (44) | 33 (70) | 38 (55) |
|     Floor of Mouth | 57 (21) | 10 (21) | 4 (6) |
|     Other | 60 (22) | 4 (9) | 15 (22) |
| WHO grade, *n* (%) | | | |
|     Mild | 91 (34) | 6 (13) | 35 (51) |
|     Moderate | 104 (39) | 25 (53) | 23 (33) |
|     Severe | 75 (28) | 16 (34) | 11 (16) |
| Binary grade, *n* (%) | | | |
|     Low-risk | 169 (63) | 7 (15) | N/A |
|     High-risk | 101 (37) | 40 (85) | N/A |
| Transformation, *n* (%) | 57 (21) | 30 (64) | 15 (22) |
| Median Follow-up Time, *months* (IQR) | 95 (57 – 108) | 41 (21 – 75) | 43 (27 – 64) |
| Scanner, *n* (%) | | | |
|     Aperio CS2 | 173 (64) | 0 (0) | 0 (0) |
|     Hamamatsu NanoZoomer 360 | 97 (36) | 0 (0) | 0 (0) |
|     Aperio AT2 | 0 (0) | 47 (100) | 0 (0) |
|     Pannoramic 250 | 0 (0) | 0 (0) | 69 (100) |

*Note. Malignancy transformation is used as the survival endpoint.*
*\*Sex not available for one participant in the Birmingham cohort.*

**Analytical Workflow**

To predict the OED risk score, we implemented a multi-step pipeline (Figure 1). First, a deep learning model was trained to automatically segment the epithelium and nuclei. This model was then used for inference on all slides. For the downstream analysis, the slides were tessellated into smaller tiles, and tile-level features were generated based on the nuclear segmentations (in tiles with ≥ 50% epithelium). These features were used to train a shallow neural network for slide-level prediction. The algorithm was internally validated on the Sheffield cohort, and subsequently validated on the two external cohorts, ensuring fully automated and unbiased testing.

*Layer and Nuclear Segmentation*

To generate layer and nuclear segmentation for each WSI in our cohort, we trained/tested HoVer-Net+ on the internal Sheffield cohort, using the ground-truth annotations. HoVer-Net+ is an encoder-decoder based CNN that simultaneously segments and classifies nuclear instances, and semantically segments the epithelial layers[18]. We used this model to semantically segment the intra-epithelial layers (e.g. basal, epithelial and keratin) and other tissue (e.g. connective tissue), whilst also segmenting and classifying nuclear instances as epithelial or other nuclei. We trained HoVer-Net+ using a multi-stage approach, based on the layer segmentations of 56 cases/controls and the nuclear segmentation masks of 24 cases/controls. The model was then tested on the layer segmentation of 12 cases/controls and the nuclear segmentations of 6 cases/controls. HoVer-Net+ takes 256×256 patches at 20× magnification (0.50 mpp), and produces nuclear instance segmentation/classification maps, and semantic segmentations of the epithelial layers. The training involved two phases, with the decoder branches trained for 20 epochs in phase one, and all branches trained for 30 epochs in phase two. A batch size of 8 and 4 on each GPU were used across these phases, respectively. The Adam optimiser was used with a learning rate that decayed initially from $10^{-4}$ to $10^{-5}$ after 10 epochs in each phase. Data augmentations such as flip, rotation, blur, and colour perturbation were applied during training. We also tested the effect of stain augmentation using the TIAToolbox[32] implementation of the Macenko method[37] that has been shown to effectively counter scanner-induced domain-shifts to make our model more generalisable[38–40]. For detailed information on model training, please refer to the **Supplementary Materials**.

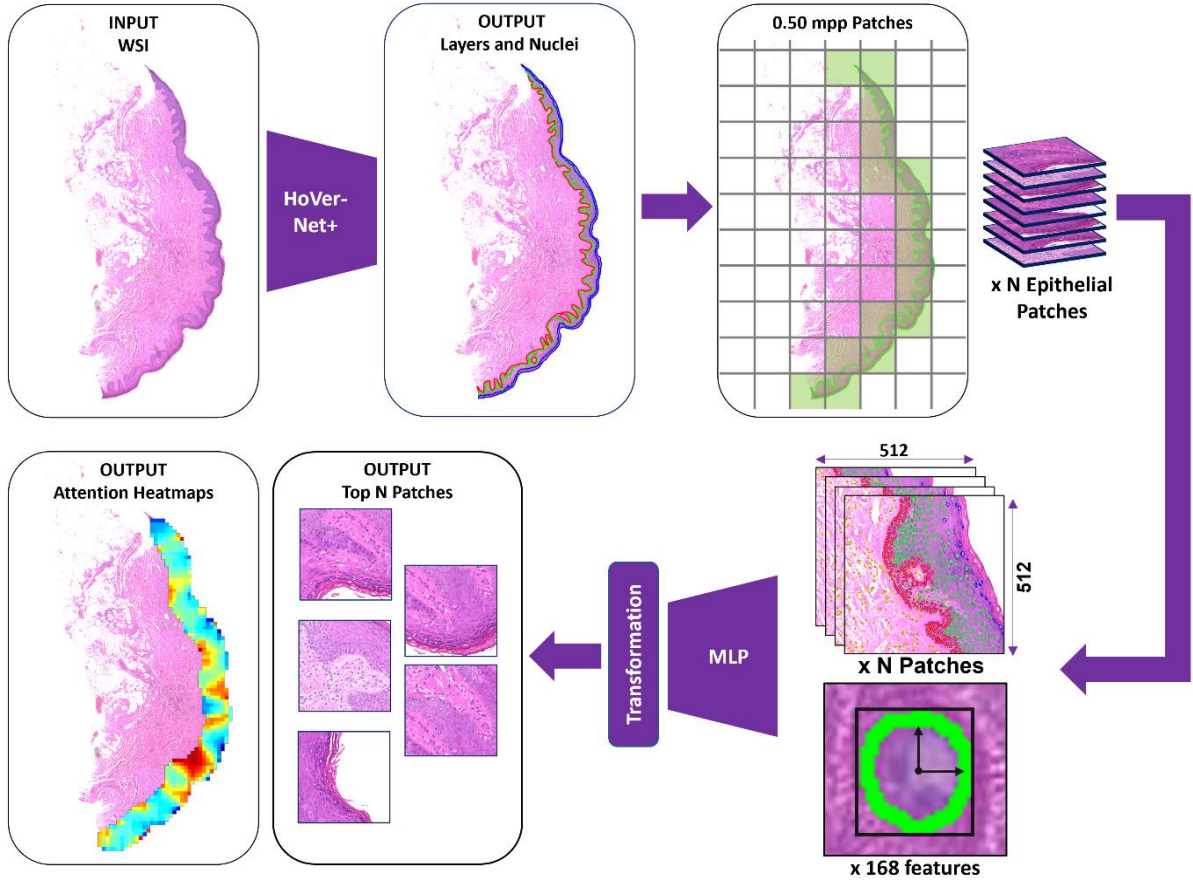

**Figure 1**
*Overview of the proposed algorithm for OMTscore, a digital score of the risk of malignant transformation of OED lesions. Here, N is the number of epithelial patches from the WSI.*

*Slide-level Transformation Prediction*

After segmentation, each WSI was tessellated into smaller 512×512 tiles (20× magnification, 0.50 mpp) with 50% overlap. We then generated tile-level features for use in a weakly supervised model for transformation prediction. For each tile, we calculated 104 morphological and 64 spatial features (see **Supplementary Materials** for further details). For slide-level prediction, a multi-layer perceptron (MLP) was trained using the iterative draw-and-rank (IDaRS) method introduced by Bilal *et al.* (2021)[41] with our tile-level features. The MLP had three layers with 168 nodes in the input layer, 64 nodes in the hidden layer, and 2 nodes in the output layer. It used a leaky ReLU activation function and dropout (0.2) after the hidden layer. The MLP models were trained with a symmetric cross-entropy loss function and the Adam optimiser. This loss function was chosen as it has been shown previously to help overcome errors associated with weak labels[41,42]. IDARS sampling was performed with parameter values of $k = 5$ for top predictive patches and $r = 45$ random patches, using a batch size of 256. Models were trained for 100 epochs and evaluated through five-fold cross-validation (repeated 3 times, with random seeds) for internal validation. We then performed external validation by combining the entire Sheffield cohort and using it as a discovery cohort for training the model, whilst validating the model on the combined Birmingham-Belfast cohort (repeated 3 times, with random seeds).

*Statistical Analyses*

Survival analyses were conducted to assess the prognostic significance of the *OMTscore*, manually-assigned WHO and binary grades, in predicting transformation-free survival. The *OMTscore* indicated whether the algorithm predicted the case to transform (high-risk) or not (low-risk). Kaplan-Meier curves[43] were generated using the Python *lifelines* package, and long-rank tests were used to determine the statistical significance of the grade stratification (OMT, WHO and binary grades). Additionally, a multivariate Cox proportional hazards model was employed, incorporating sex, age, lesion site, and grade, to predict transformation-free survival. The purpose of this analysis was to validate the prognostic significance of the predicted *OMTscore* relative to other clinical variables. This analysis was performed on both the internal and external cohorts. Transformations were right-censored at five years across these analyses to ensure consistency between internal and external cohorts.

*Feature Analysis*

In the post-hoc analysis of our internal validation cohort, we focused on the nuclear count features within the top five predicted patches of correctly predicted positive slides (true positives) and compared them to the bottom five predicted patches of correctly predicted negative slides (true negatives) within the testing subsets. Two-tailed t-tests were performed with multiple comparison correction (false discovery rate, FDR) to determine the statistical significance of any observed differences. We conducted three comparative analyses of the top/bottom predicted patches: 1) nuclei within the entire patch (other, basal, epithelial, keratin), 2) nuclei within the epithelium (other, basal, epithelial, keratin), and 3) nuclei within the connective tissue surrounding the patch (e.g., peri-epithelial "other" nuclei). In addition, we analysed the tissue type ratios within these top/bottom predicted patches.

*Evaluation Metrics*

We evaluated the layer segmentation using the F1-score aggregated over all image patches. For nuclear instance segmentation, we assessed the Panoptic Quality (PQ), detection quality (DQ, or F1-score) and segmentation quality (SQ). Additionally, we report the Dice score comparing segmented nuclei against the background, and the aggregated Jaccard Index (AJI)[44]. We also calculate the average values over all images for: F1-score for detection ($F_d$, all nuclear types) and F1-score for classification ($F_c$) for each nucleus type (e.g. $F_c^b$ for basal epithelial nuclei, $F_c^e$ for epithelial nuclei, and $F_c^o$ for other nuclei). Detailed descriptions of these metrics can be found in Graham et al. (2019)[17]. When evaluating the models' performance in predicting transformation, we calculated the average F1-score and AUROC across all slides.

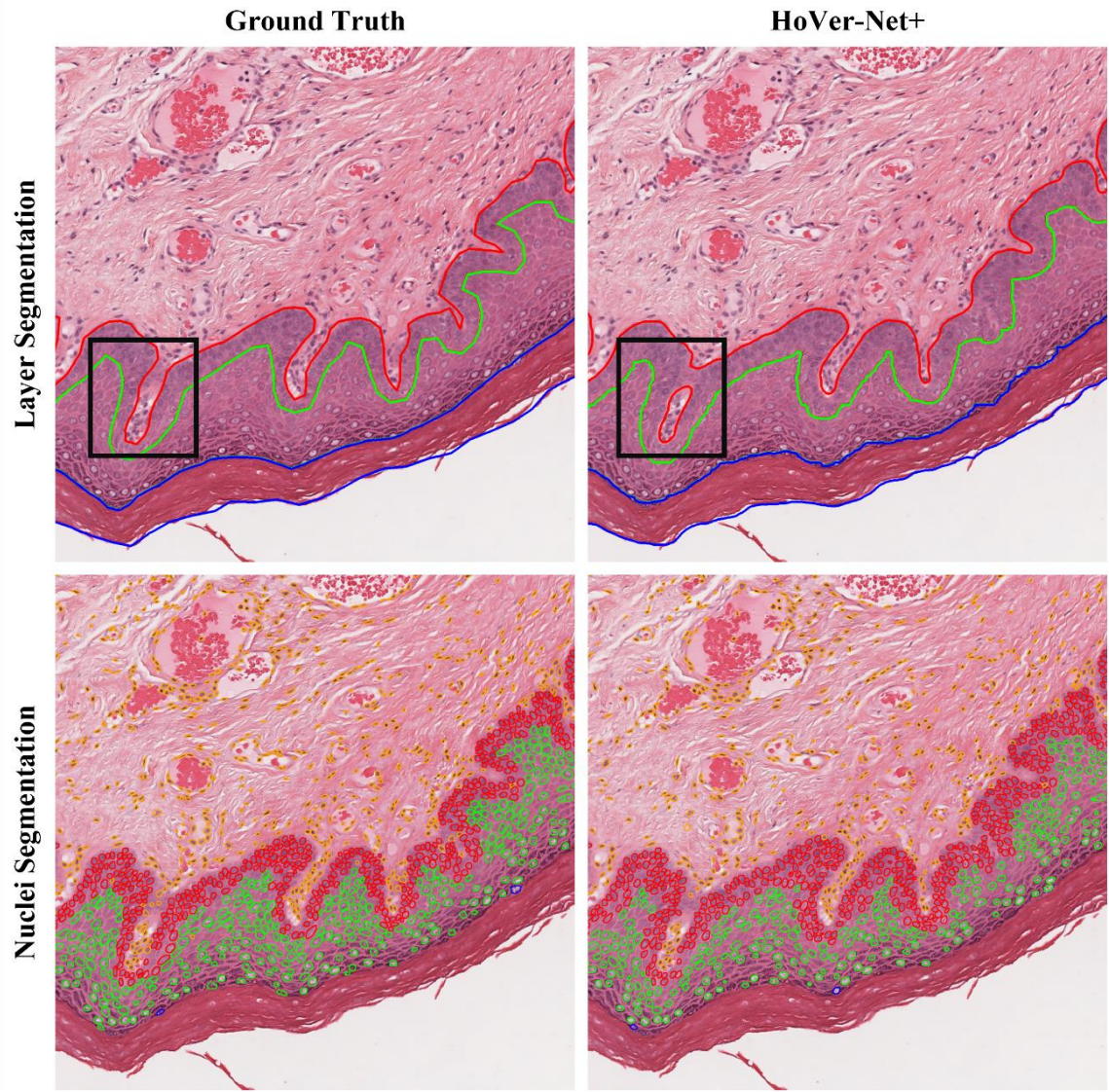

**Figure 2**
*A comparison of the ground truth layer/nuclei segmentations and HoVer-Net+. The red, blue and green lines and contours represent the layers and nuclei in the basal, epithelial and keratin layers. The orange contours represent other nuclei outside the three layers. The black box shows an area where HoVer-Net+ has correctly annotated basal layer which the pathologist had missed.*

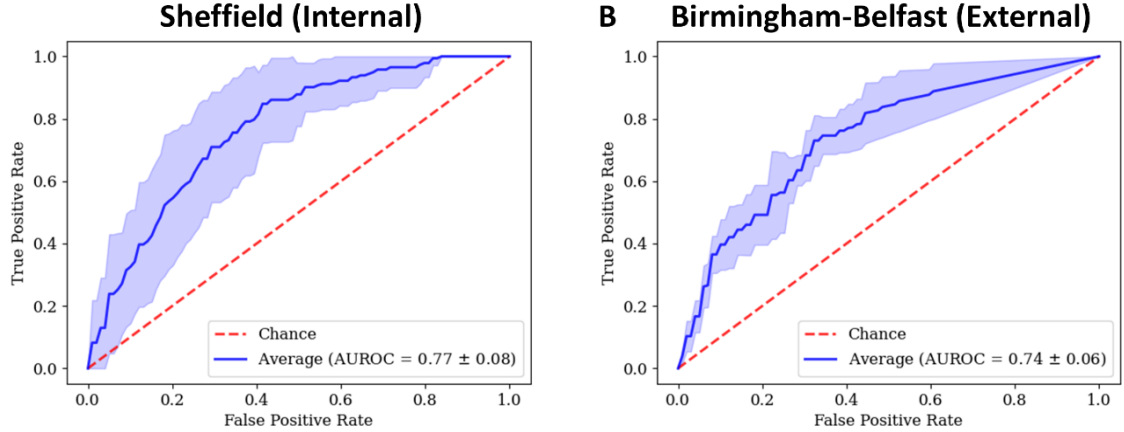

**Figure 3**

*ROC plots for predicting malignant transformation with A) internal validation on Sheffield, and b) external validation on the Birmingham-Belfast datasets by the MLP.*

## Results

**Layer and Nuclei Segmentation**

Following training, our model achieved the following average F1-scores for the semantic segmentation of background (F1 = 0.876), other tissue (F1 = 0.838), basal epithelium (F1 = 0.740), epithelium (F1 = 0.868), and keratin (F1 = 0.813), yielding a mean F1-score of 0.813. In terms of nuclear instance segmentation and classification, our model achieved a Dice score of 0.691, while the other metric values are as follows: AJI = 0.617, DQ = 0.739, SQ = 0.689, PQ = 506, $F^d$ = 0.822, $F_c^o$ = 0.722, $F_c^b$ = 0.607, and $F_c^e$ = 0.657. For a visual comparison between HoVer-Net+ results and ground-truth annotations, refer to Figure 2. Overall, we deemed these results satisfactory and thus used the trained HoVer-Net+ model for inference on cases from all three cohorts. To ensure the quality of segmentation on the external cohorts, we conducted a visual assessment of HoVer-Net+'s performance and excluded 24 cases from the Birmingham-Belfast cohorts due to poor segmentation results, resulting in a total of 92 OED cases with 39 transformations.

**Slide-level Transformation Prediction**

In these experiments, we compared the performance of our model, trained with patch-level morphological/spatial features, to the pathologist grades (see Table 2). On internal validation, our model attained competitive results with an AUROC of 0.77, outperforming both the WHO grade (AUROC = 0.58) and the binary grade (AUROC = 0.71). For external validation on the Birmingham-Belfast cohort, our model again achieved superior results (AUROC = 0.74) compared to both the WHO and binary grades. The ROC curves for our proposed model are shown in Figure 3. These results demonstrate the superiority of the proposed model for predicting OED transformation when compared to the manual pathologist grading.

**Table 2**

*Slide-level mean (standard deviation) results for transformation prediction.*

| | | Internal Validation | | External Validation | | | | | |
| | | Sheffield | | Birmingham | | Belfast | | Combined | |
| Model | Features | F1-score | AUROC | F1-score | AUROC | F1-score | AUROC | F1-score | AUROC |
| --- | --- | --- | --- | --- | --- | --- | --- | --- | --- |
| **MLP** | **Morph/Spatial** | **0.57 (0.08)** | **0.77 (0.08)** | **0.64 (0.01)** | **0.78 (0.02)** | 0.73 (0.09) | **0.71 (0.06)** | **0.70 (0.05)** | **0.74 (0.08)** |
| Binary Grade | Low- vs High-Risk | 0.51 (0.08) | 0.71 (0.06) | - | - | **0.74** | 0.52 | - | - |
| WHO Grade | Mild/Mod. vs Severe | 0.34 (0.16) | 0.58 (0.11) | 0.44 | 0.64 | 0.35 | 0.40 | 0.36 | 0.54 |
| WHO Grade | Mild vs Mod./Severe | 0.46 (0.08) | 0.68 (0.05) | 0.61 | 0.75 | 0.73 | 0.49 | 0.64 | 0.64 |

**Table 3**

*Cox Proportional Hazard Model output for malignant transformation based on the OMTscore, and other clinical variables.*

| | Internal Validation – Sheffield ($n = 270$) | | | | External Validation – Combined ($n = 92$) | | | |
| | *p* | HR | Lower 95% HR | Upper 95% HR | *p* | HR | Lower 95% HR | Upper 95% HR |
| --- | --- | --- | --- | --- | --- | --- | --- | --- |
| ***OMTscore*** | **< 0.0001** | **8.64** | **3.18** | **27.13** | **0.01** | **7.34** | **0.85** | **125.40** |
| **WHO Grade** | **< 0.0001** | **2.49** | **1.52** | **4.00** | 0.92 | 1.16 | 0.60 | 2.08 |
| Age | 1.00 | 1.03 | 0.71 | 1.48 | 0.38 | 1.02 | 0.99 | 1.05 |
| Sex | 1.00 | 1.09 | 0.57 | 2.19 | 0.62 | 1.44 | 0.63 | 3.16 |
| Site | 1.00 | 1.00 | 0.98 | 1.03 | 0.92 | 1.20 | 0.73 | 1.94 |

*HR = Hazard Ratio.*

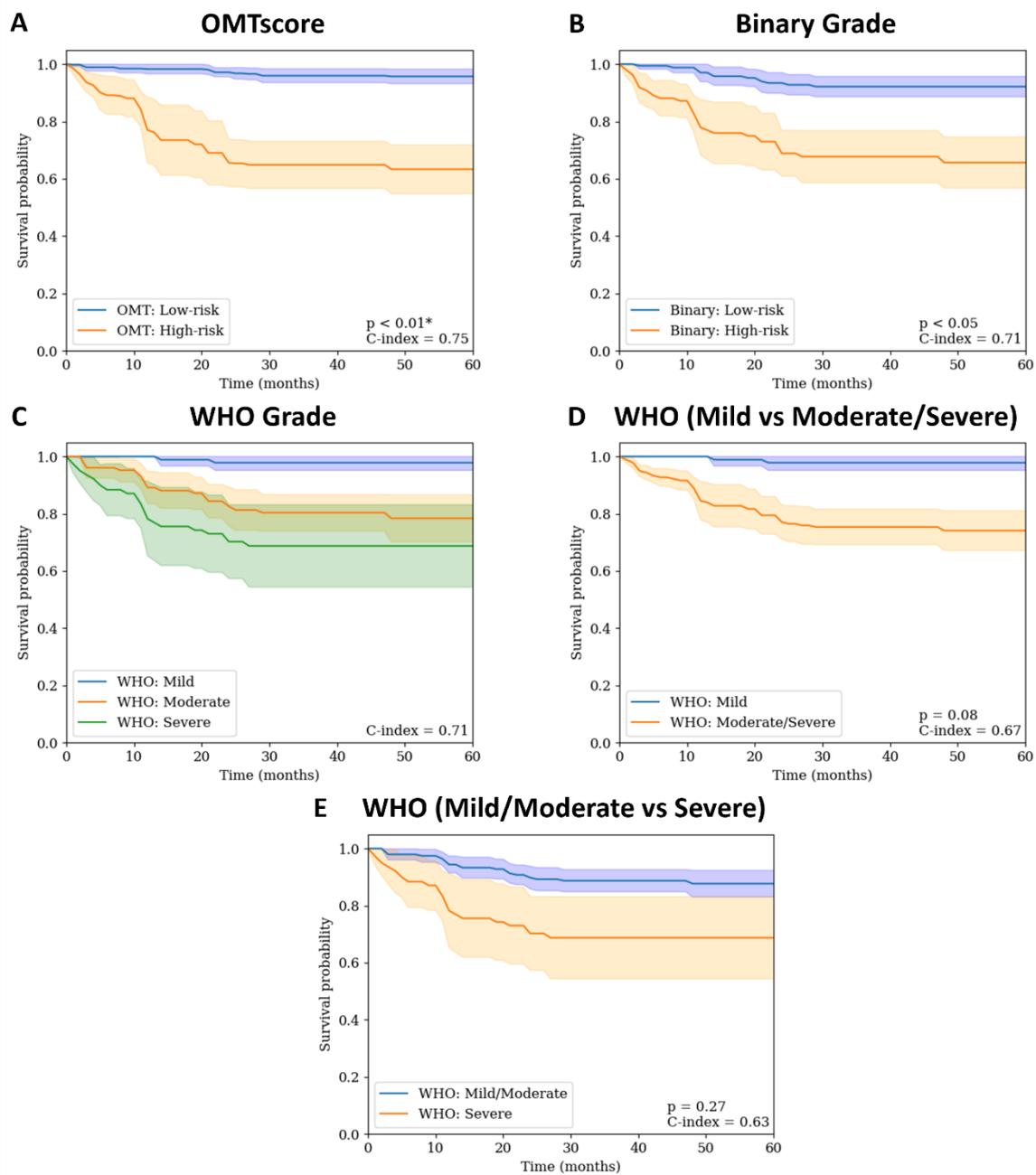

**Figure 4**
*Kaplan-Meier survival curves for transformation-free survival based of the predictions from A) MLP OMTscore, B) binary grade, C) WHO grade, D) WHO grade (mild vs. moderate/severe), and E) WHO grade (mild/moderate vs. severe).*

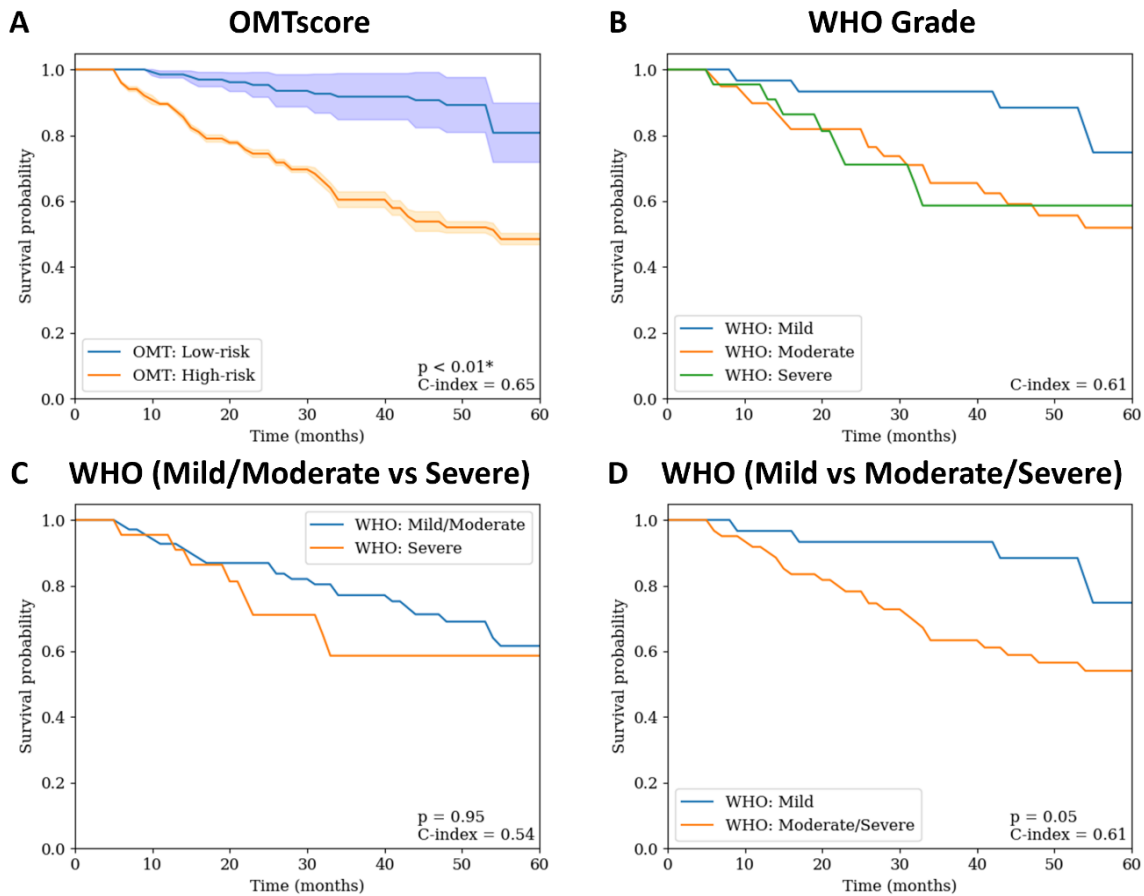

**Figure 5**
*Kaplan-Meier survival curves for transformation-free survival based of the predictions from A) MLP OMT-grade, B) WHO grade, C) WHO grade (mild/moderate vs. severe), and D) WHO grade (mild vs. moderate/severe).*

**Survival Analysis**

We plotted Kaplan-Meier (KM) curves to assess the prognostic value of our *OMTscore* (Figure 4A) and binary/WHO grades (Figure 4B-E) on the internal cohort. The *OMTscore* demonstrated a clear separation between the low- and high-risk cases ($p < 0.01$) outperforming the binary grade ($p < 0.05$) and WHO grade. The mean C-index of our model (C-index = 0.75) was higher than that of the binary (C-index = 0.71) and WHO grades. Results from the Cox proportional hazard model (Table 3) showed that both the WHO grade and *OMTscore* were statistically significant, with the *OMTscore* exhibiting the highest hazard ratio (HR), indicating better prognostic utility.

For external validation, KM survival curves were presented for the Birmingham-Belfast cohort (Figure 5). Only the *OMTscore* exhibited statistically significant differences in KM curves ($p < 0.01$) according to a log-rank test. The *OMTscore* achieved an improved C-index of 0.65 compared to the WHO grade's C-index of 0.61. Results from the multivariate Cox PH models (Table 3) again showed statistical significance for the *OMTscore* with a high hazard ratio ($p = 0.01$, HR = 7.34), highlighting its prognostic utility over the other clinical variables and the WHO grade.

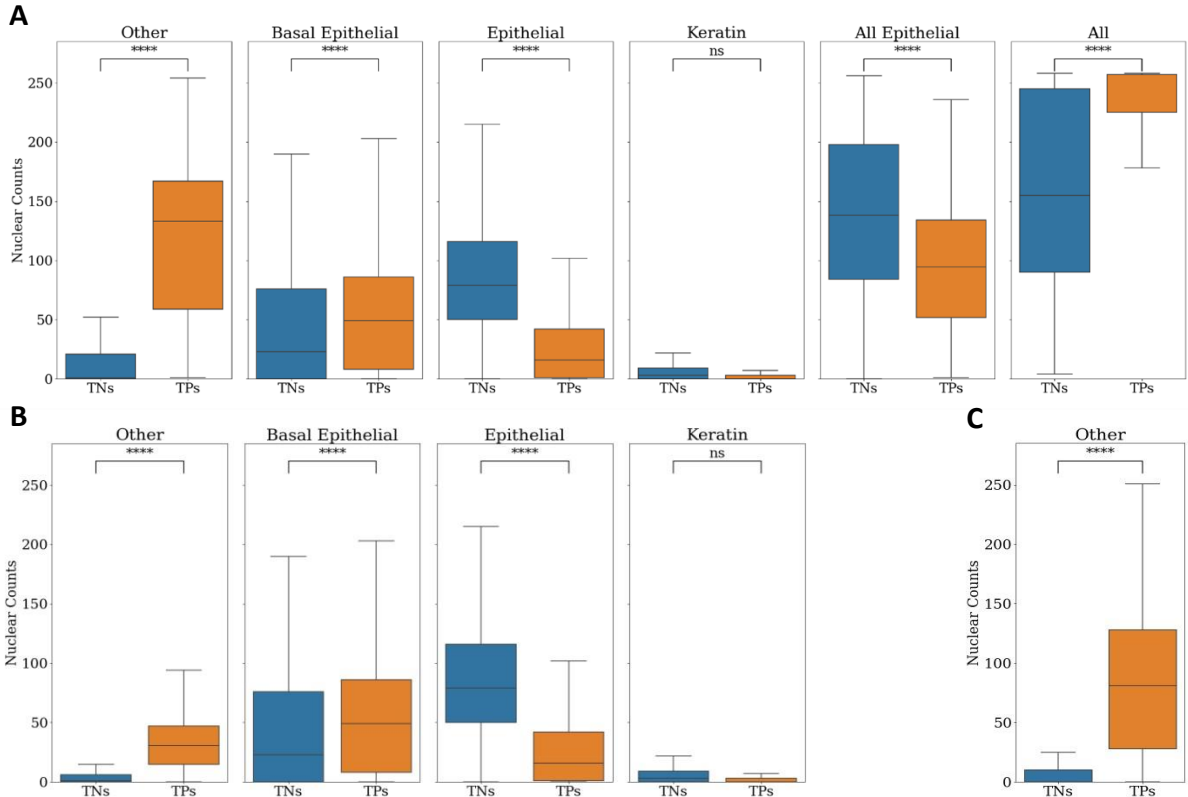

**Figure 7**
*Boxplots showing the distribution of nuclear counts within A) the entire patch, B) the epithelium alone, and C) the connective tissue alone, of the top five predicted patches from true positive cases, and the bottom five patches from true negatives.*

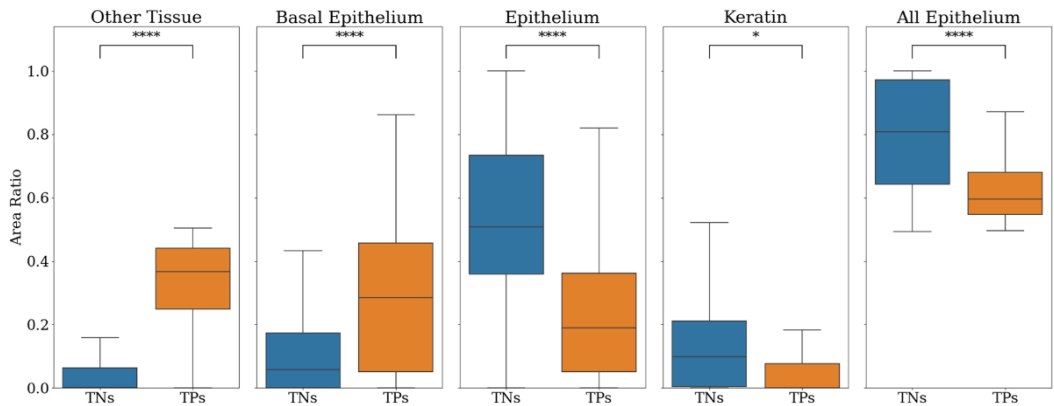

**Figure 6**
*Boxplots showing the distribution of region areas within the top five predicted patches from true positive cases, and the bottom five patches from true negatives.*

**Feature Analysis**

We analysed the most important features used by our model by comparing the top five predictive patches in true positive cases to the bottom predicted patches in true negative cases (see the **Supplementary Material** for a random selection of patches). Patch-level nuclear counts revealed higher cellularity in TP patches compared to TNs (Cohen's $d = 0.69$, $p < 0.0001$; see Figure 7A), primarily driven by "other nuclei" in TPs ($d = 2.25$, $p < 0.0001$). In contrast, there were slightly more epithelial cells in TNs (labelled as "All Epithelial" in Figure 7 A; $d = 0.10$, $p < 0.0001$). The higher number of "other nuclei" in these patches may be due to an increase in the number of epithelial-infiltrating lymphocytes and peri-epithelial lymphocytes. When focussing on the nuclear counts within the epithelial region of the patch alone (Figure 7B), significant differences were found in the number of "other" nuclei within the epithelium ($d = 2.29$, $p < 0.0001$), confirming the disparity in the number of epithelial-infiltrating lymphocytes. Additionally, there were more epithelial nuclei within the epithelial layer in TNs ($d = 1.31$, $p < 0.0001$), while more basal epithelial nuclei were observed in TPs ($d = 0.24$, $p < 0.0001$). No difference was found in the number of keratin nuclei between classes. Lastly, Figure 7C illustrates a larger number of "other" nuclei within the connective tissue of TPs compared to TNs ($d = 1.73$, $p < 0.0001$), confirming the increased presence of peri-epithelial lymphocytes in TPs.

When analysing the distributions of tissue types within patches (Figure 6), we found that TP patches had a higher ratio of connective tissue (presumed from "other" tissue) compared to TNs ($d = 2.21$, $p < 0.0001$). This is consistent with the prior nuclear analysis showing more "other" nuclei in TP patches. Additionally, TP patches often had more basal tissue ($d = 1.02$, $p < 0.0001$), but less epithelial tissue ($d = 1.33$, $p < 0.0001$), compared to TNs. Interestingly, TNs had significantly more surface keratin compared to TPs ($d = 0.10$, $p < 0.02$).

Overall, these results suggest that cases that did not transform tend to have more epithelial cells, while cases that did transform have more *basal* epithelial cells. This aligns with the known manifestation of OED within the basal layer of the epithelium. Additionally, cases that transformed exhibited higher numbers of "other" nuclei both the connective tissue and the epithelium. This is indicative of higher amounts of peri-epithelial lymphocytes, as found by Bashir *et al.*[34], as well as higher numbers of epithelial-infiltrating lymphocytes, a potentially novel finding of this analysis. TN patches primarily focussed on the epithelium, with little emphasis on the connective tissue, while TP patches showed a focus on the basal layer and connective tissue.

The heatmaps produced by our model were inspected by a pathologist (SAK). They revealed prognostic areas with obvious or high grades of dysplasia, and a significant presence of immune cells within and around the epithelium. An example heatmap of a mild OED case is shown in Figure 8, which was correctly predicted by our model to transform. Further examination of the hotspots indicated a focus on dysplastic areas with a prominent lymphocytic infiltrate within the epithelium and peri-epithelial lymphocytes, consistent with the findings from our nuclear analyses.

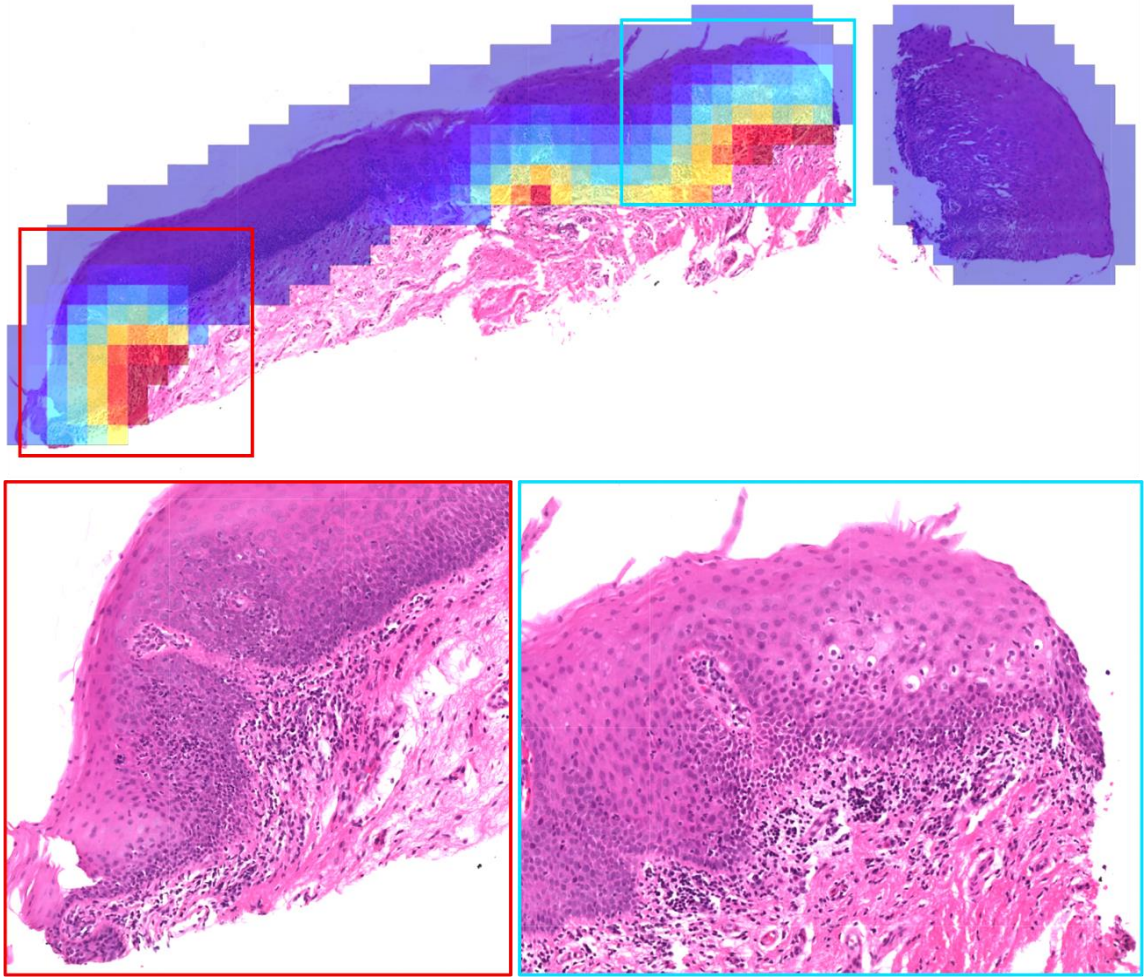

**Figure 8**
*A mild OED WSI with the model prediction heatmap overlaid, where the model correctly predicted this case to transform to malignancy in the internal validation cohort.*

**Discussion**

We proposed a novel risk score using interpretable morphological and spatial features that outperforms manual grades in predicting OED transformation. Our approach involved first introducing a new model for simultaneous segmentation of intra-epithelial layers and nuclei in H&E WSIs. We then generated patch-level morphological/spatial features, resembling cytological features used by pathologists for OED grading (e.g. anisonucleosis, nuclear pleomorphism). These features were fed into a shallow neural network, yielding high predictive performance for OED transformation. The generated *OMTscore* showed superior prognostic significance compared to both the binary and WHO grading systems, with a clear distinction between low- and high- risk cases ($p < 0.01$).

The morphological/spatial features used in our model, based on shape and size differences in different nuclear types, are theoretically domain-agnostic, contributing to our model's robust performance on external cohorts. We argue that incorporating such features in favour of deep features makes our models more interpretable and explainable. Our post-hoc analyses highlighted the importance of peri-epithelial nuclei (in line with Bashir *et al.*[34]) and a higher density of epithelial-infiltrating lymphocytes in predicting OED transformation. The latter finding is interesting, as within oral cancer, lesions exhibiting more intra-epithelial lymphocytes, have typically amounted a higher immune response and therefore often have a better outcome. Within oral dysplasia, we therefore suggest the opposite, that a higher immune response may instead be indicative that the lesion will transform to cancer. This is a novel finding warranting further investigation.

The binary grading system showed better prognostic value than the WHO grading system, effectively separating low- and high-risk cases. Our risk score outperformed pathologist/histological grades, achieving a higher performance (AUROC = 0.74) on the external cohorts compared to both the WHO and binary grading systems. This demonstrates the superior prognostic utility of our *OMTscore*.

Our study has some limitations. We collected 193 OED cases for the development and internal validation and 92 cases for external validation. Although this may be the largest known OED dataset with clinical outcomes for computational pathology, the sample size is still relatively small. Exclusion of some cases (a total of 24) from the external cohort was a result of the HoVer-Net+ model not producing satisfactory segmentation results and is not reflective of the quality of the proposed transformation prediction pipeline. The HoVer-Net+ model requires further generalisability, potentially with the addition of more cases from different centres. For completeness, we include additional analysis in the **Supplementary Material**, when incorporating these cases into the final pipeline. Future work should focus on improving the generalisability of the HoVer-Net+ model to unseen datasets and exploring additional pathologist-derived patch-level features, such as hyperchromatism and mitoses, that may better predict OED transformation,. Further inspiration can be drawn from Mahmood et al. (2022) for developing such features.

In summary, our study has introduced a novel automated pipeline for the predicting OED transformation using a state-of-the-art deep learning framework and patch-level morphological/spatial features. Our results demonstrate the strong prognostic significance and generalisability of our model compared to manual grades on internal and external cohorts. This has significant clinical implications for patient management, offering a more accurate prediction method. Our study paves the way for future research and the potential to enhance patient outcomes through early detection and intervention. However, further investigations are required to identify additional slide-level features and validate the model on larger external cohorts with longer follow-up periods.

## Acknowledgements

This work was supported by a Cancer Research UK Early Detection Project Grant, as part of the ANTICIPATE study (grant no. C63489/A29674). SAK and NMR are partly funded by the CRUK grant. HM is funded by the National Institute for Health Research (award no. NIHR300904). RMSB is funded by the Chancellor Scholarship from University of Warwick. The authors express their sincere gratitude to Professor Paul Speight (PMS), Professor Paula Farthing (PMF), and Dr Daniel Brierley (DJB) for their valuable contribution in providing the initial histological diagnosis.

## Disclosure/Conflict of Interest

The authors declare the following competing interest: NMR is the co-founder, Director and a shareholder of Histofy Ltd. SAK is a shareholder of Histofy Ltd. All other authors have no competing interests to declare.

# References

1. Johnson, D. E. *et al.* Head and neck squamous cell carcinoma. *Nat. Rev. Dis. Prim.* **6**, (2020).

2. European Cancer Patient Coalition. European Cancer Patient Coalition: Head & Neck Cancers. https://ecpc.org/news-events/head-neck-cancer-make-sense-campaign/.

3. Speight, P. M., Khurram, S. A. & Kujan, O. Oral potentially malignant disorders: risk of progression to malignancy. *Oral Surg. Oral Med. Oral Pathol. Oral Radiol.* **125**, 612–627 (2018).

4. Ranganathan, K. & Kavitha, L. Oral epithelial dysplasia: Classifications and clinical relevance in risk assessment of oral potentially malignant disorders. *J. Oral Maxillofac. Pathol.* **23**, 19–27 (2019).

5. Rock, L. D. *et al.* Characterization of epithelial oral dysplasia in non-smokers: First steps towards precision medicine. *Oral Oncol.* **78**, 119–125 (2018).

6. Nankivell, P. & Mehanna, H. Oral dysplasia: Biomarkers, treatment, and follow-up. *Curr. Oncol. Rep.* **13**, 145–152 (2011).

7. Mehanna, H. M., Rattay, T., Smith, J. & McConkey, C. C. Treatment and follow-up of oral dysplasia — A systematic review and meta-analysis. *Head Neck* **31**, 1600–9 (2009).

8. Takata, T. & Slootweg, P. Tumours of the oral cavity and mobile tongue: epithelial precursor lesions. in *WHO Classification of Head and Neck Tumours* (eds. El-Naggar, A., Chan, J., Grandis, J., Takata, T. & Slootweg, P.) (2017).

9. Nankivell, P. *et al.* The binary oral dysplasia grading system: validity testing and suggested improvement. *Oral Surg. Oral Med. Oral Pathol. Oral Radiol.* **115**, 87–94 (2013).

10. Lecun, Y., Bengio, Y. & Hinton, G. Deep learning. *Nature* **521**, 436–444 (2015).

11. Litjens, G. *et al.* Deep learning as a tool for increased accuracy and efficiency of histopathological diagnosis. *Sci. Rep.* **6**, 1–11 (2016).

12. Madabhushi, A. & Lee, G. Image analysis and machine learning in digital pathology: Challenges and opportunities. *Med. Image Anal.* **33**, 170–175 (2016).

13. Litjens, G. *et al.* A survey on deep learning in medical image analysis. *Med. Image Anal.* **42**, 60–88 (2017).

14. Ghafoorian, M. *et al.* Location Sensitive Deep Convolutional Neural Networks for Segmentation of White Matter Hyperintensities. *Sci. Rep.* **7**, 5110 (2017).

15. Liu, J. *et al.* Applications of deep learning to MRI images: A survey. *Big Data Min. Anal.* **1**, 1–18 (2018).

16. Shen, D., Wu, G. & Suk, H.-I. Deep Learning in Medical Image Analysis. *Annu. Rev. Biomed. Eng.* **19**, 221–248 (2017).

17. Graham, S. *et al.* Hover-Net: Simultaneous segmentation and classification of nuclei in multi-tissue histology images. *Med. Image Anal.* **58**, 101563 (2019).

18. Shephard, A. J. *et al.* Simultaneous Nuclear Instance and Layer Segmentation in Oral Epithelial Dysplasia. *Proc. IEEE/CVF Int. Conf. Comput. Vis. Work.* **October**, 552–561

(2021).

19. Alemi Koohbanani, N., Jahanifar, M., Zamani Tajadin, N. & Rajpoot, N. NuClick: A deep learning framework for interactive segmentation of microscopic images. *Med. Image Anal.* **65**, (2020).

20. Raza, S. E. A. *et al.* Micro-Net: A unified model for segmentation of various objects in microscopy images. *Med. Image Anal.* **52**, 160–173 (2019).

21. Azarmehr, N., Shephard, A., Mahmood, H., Rajpoot, N. & Khurram, S. A. A Neural Architecture Search Based Framework for Segmentation of Epithelium, Nuclei and Oral Epithelial Dysplasia Grading. in *Annual Conference on Medical Image Understanding and Analysis MIUA 2022* vol. 13413 357–370 (Springer International Publishing).

22. Bulten, W. *et al.* Epithelium segmentation using deep learning in H&E-stained prostate specimens with immunohistochemistry as reference standard. *Sci. Rep.* **9**, 1–10 (2019).

23. Bashir, R. M. S. *et al.* Automated grade classification of oral epithelial dysplasia using morphometric analysis of histology images. in *Medical Imaging 2020: Digital Pathology* vol. 11320 (International Society for Optics and Photonics, 2020).

24. Lu, M. Y. *et al.* Data-efficient and weakly supervised computational pathology on whole-slide images. *Nat. Biomed. Eng.* **5**, 555–570 (2021).

25. Zhou, Y. *et al.* CGC-Net: Cell graph convolutional network for grading of colorectal cancer histology images. *arXiv* (2019).

26. Shephard, A. *et al.* A Fully Automated Multi-Scale Pipeline for Oral Epithelial Dysplasia Grading and Outcome Prediction. *Med. Imaging with Deep Learn.* 1–3 (2022).

27. Lu, W. *et al.* SlideGraph+: Whole slide image level graphs to predict HER2 status in breast cancer. *Med. Image Anal.* **80**, 102486 (2022).

28. Kather, J. N. *et al.* Deep learning can predict microsatellite instability directly from histology in gastrointestinal cancer. *Nat. Med.* **25**, 1054–1056 (2019).

29. Campanella, G. *et al.* Clinical-grade computational pathology using weakly supervised deep learning on whole slide images. *Nat. Med.* **25**, 1301–1309 (2019).

30. Ilse, M., Tomczak, J. M. & Welling, M. Attention-based deep multiple instance learning. *35th Int. Conf. Mach. Learn. ICML 2018* **5**, 3376–3391 (2018).

31. Bilal, M. *et al.* An Aggregation of Aggregation Methods in Computational Pathology. *Med. Image Anal.* 102885 (2022) doi:10.1016/J.MEDIA.2023.102885.

32. Pocock, J. *et al.* TIAToolbox as an end-to-end library for advanced tissue image analytics. *Commun. Med.* **2**, 120 (2022).

33. Mahmood, H. *et al.* Prediction of malignant transformation and recurrence of oral epithelial dysplasia using architectural and cytological feature specific prognostic models. *Mod. Pathol.* **35**, 1151–1159 (2022).

34. Bashir, R. M. S. *et al.* A digital score of peri-epithelial lymphocytic activity predicts malignant transformation in oral epithelial dysplasia. *J. Pathol.* (2023) doi:10.1002/path.6094.

35. Gutman, D. A. *et al.* The digital slide archive: A software platform for management, integration, and analysis of histology for cancer research. *Cancer Res.* **77**, e75–e78 (2017).

36. Jahanifar, M., Koohbanani, N. A. & Rajpoot, N. NuClick: From Clicks in the Nuclei to Nuclear Boundaries. *arXiv* (2019).

37. Macenko, M. *et al.* A method for normalizing histology slides for quantitative analysis. *Proc. - 2009 IEEE Int. Symp. Biomed. Imaging From Nano to Macro, ISBI 2009* 1107–1110 (2009) doi:10.1109/ISBI.2009.5193250.

38. Aubreville, M. *et al.* Mitosis domain generalization in histopathology images — The MIDOG challenge. *Medical Image Analysis* vol. 84 102699 (2023).

39. Jahanifar, M. *et al.* Stain-Robust Mitotic Figure Detection for the Mitosis Domain Generalization Challenge. *arXiv* 3–5 (2021).

40. Jahanifar, M., Shephard, A., Zamanitajeddin, N., Raza, S. E. A. & Rajpoot, N. Stain-Robust Mitotic Figure Detection for MIDOG 2022 Challenge. 1–3 (2022).

41. Bilal, M. *et al.* Development and validation of a weakly supervised deep learning framework to predict the status of molecular pathways and key mutations in colorectal cancer from routine histology images: a retrospective study. *Lancet Digit. Heal.* **3**, e763–e772 (2021).

42. Wang, Y. *et al.* Symmetric Cross Entropy for Robust Learning with Noisy Labels.

43. Kaplan, E. L. & Meier, P. Nonparametric Estimation from Incomplete Observations. *J. Am. Stat. Assoc.* **53**, 457–481 (1958).

44. Kumar, N. *et al.* A Dataset and a Technique for Generalized Nuclear Segmentation for Computational Pathology. *IEEE Trans. Med. Imaging* **36**, 1550–1560 (2017).

# Supplementary Materials

**Data**

*Sheffield Inclusion Criteria*

WSIs were collected into this cohort based on the following criteria: a histological diagnosis of oral epithelial dysplasia (OED), sufficient availability of tissue (i.e. excluding tangentially cut sections, tissue with artefacts, verrucous lesions) and at least 5 years of follow-up data from the initial diagnosis (including treatment and transformation information). To ensure diagnostic consistency, all cases were independently evaluated by at least two certified/consultant pathologists (PMS, PMF, DJB), who provided an initial diagnosis based on the WHO grading system (2008-2016). To confirm the WHO (2017) grade and assign binary grades, the cases were blindly re-evaluated by an Oral & Maxillofacial Pathologist (SAK) and an Oral Surgeon with a specialist interest and expertise in OED analysis (HM). The interobserver disagreement between the two pathologists was assessed using Cohen's kappa score, which resulted in a value of 0.854. This indicates a high level of agreement between the two pathologists. Cases with disagreement were resolved through discussion within the team.

**Layer and Nuclei Segmentation Pipeline**

*Materials*

For training our segmentation models, an expert oral and maxillofacial pathologist (SAK) exhaustively manually delineated the three layers of the epithelium (basal layer, epithelial layer, superior keratin layer) in 59 OED cases, in addition to nine controls (collected with the Aperio scanner subject to the above protocols), using our inhouse WASABI software (a customised version of HistomicsTK). We then generated tissue masks for each of the segmented WSIs via Otsu thresholding and the further removal of small holes and objects. A layer mask was then generated for each WSI by combining the layer segmentations with the tissue mask.

The manual segmentation of individual nuclei within WSIs is laborious and subject to inter/intra-rater variability. Thus nuclear instance masks were only generated for a subset of cases. 30 regions of interest (ROIs) were created (one per case), and the pathologist (SAK) placed a point annotation on each nucleus within each of the ROIs. Each nucleus was labelled as either epithelial nuclei or 'other' nuclei. The point annotations were then used within the NuClick framework to generate nuclear boundaries[1,2]. NuClick is a deep learning framework that takes a raw image and a guiding signal 'click' as an input and then produces a nuclear instance boundary as an output. This method has been found previously to be superior to fully-automated approaches for generating nuclear instance segmentations, particularly in the cases of touching/overlapping nuclei[1]. Following the generation of nuclear instance masks in all 30 ROIs, a quality control step was added to ensure that all nuclei segmentations were of a high quality. The masks were then manually refined when found to be visibly incorrect. This typically occurred to nuclei in the granular layer of the epithelium. In total, this method resulted in 71,757 labelled nuclei segmentations across the 30 ROIs. These annotations were then used to train our deep learning segmentation models.

We aimed to make HoVer-Net+ as generalisable as possible; thus, in the training of our model we further split up our Sheffield dataset into smaller cohorts for testing the model generalisability. Thus of the 68 cases/controls that we had layer (and some nuclear) annotations for, we further split it up into two cohorts, cohort 1 (C1) and cohort 2 (C2; see Table S1 for breakdown), so that each cohort had images from two different scanners.

**Table S1**

*Sheffield cohorts used for training HoVer-Net+.*

|  | Layers | | Nuclei | |
| --- | --- | --- | --- | --- |
|  | Cohort 1 | Cohort 2 | Cohort 1 | Cohort 2 |
| Cases, *n* | | | | |
|    Aperio CS2 | 38 | 0 | 20 | 0 |
|    Hamamatsu NanoZoomer 360 | 0 | 21 | 0 | 10 |
| Controls, *n* | | | | |
|    Aperio CS2 | 5 | 4 | 0 | 0 |
|    Hamamatsu NanoZoomer 360 | 0 | 0 | 0 | 0 |

*Methods*

HoVer-Net+ is multi-task learning method, thus, it is assumed that the training of the encoder via different datasets/tasks will only increase its generalisability and performance[3]. To train our HoVer-Net+ model, we have taken a multi-stage, multi-cohort approach. At 20× magnification (0.50 mpp), it is logical that upon patch extraction, we will produce many more patches with layer segmentations (68 WSIs) than nuclear annotations (30 ROIs). Thus, the primary step of our training schemes will be to train the encoder and LS decoder for layer segmentation. Following this, we could then train the nuclear segmentation/classification decoders (NP/HoVer/NC branches). We tessellated our WSIs, layer masks and nuclear instance maps into smaller patches of size $256 \times 256$ (no overlap) at 20× magnification (0.50 mpp). For C1, this resulted in 1,139 patches with nuclear instance maps (and corresponding layer masks), and 24,532 patches with layer masks. For C2, this resulted in 866 patches with nuclear instance maps, and over 7,731 patches with layer masks. For training purposes, an 80/20 train/test split was used for all models. Our training scheme consists of three stages:

1. We first aim to train HoVer-Net+ for layer segmentation alone, based on all the layer segmentation patches (model: HoVer-Net+$_{LS}$).
2. We then aim to train the entire network based on patches with both layer and nuclear annotations (model: HoVer-Net+$_{LS,All}$).
3. Finally, we freeze the encoder and train the LS decoder only on the segmentation patches again (model: HoVer-Net+$_{LS,All,LS}$)

During the second stage of training the LS branch was trained based on less layer annotations than in stage one, and thus will be overfit to the smaller sample used. Thus, the third stage aims to retrain the LS decoder again with all the layer annotations available.

We trained the HoVer-Net+ models in stages one and two over two phases. In phase one, only the decoder branches were trained for 20 epochs. In phase two, all branches (including the encoder) were trained for 30 epochs. A batch size of 8 and 4 on each GPU were used across these two phases respectively. In stage three, only the second phase of training was used as the encoder was kept frozen. For all training, the Adam optimiser was used with a learning rate that decayed initially from $10^{-4}$ to $10^{-5}$ after 10 epochs in each phase. During training, we applied the following random data augmentations: flip, rotation, Gaussian blur, median blur, and colour perturbation. We additionally, tested the effect of stain augmentation during our experiments, using the TIAToolbox[4] implementation, a method that has been shown to effectively counter scanner-induced domain-shifts, in order to create more generalisable models[5–7].

Comparative experiments were performed to test HoVer-Net+ model against other state-of-the-art models for nuclear segmentation and classification. For layer segmentation, we compared HoVer-Net+ to U-Net[8], a state-of-the-art model for semantic segmentation, following stage one of layer segmentation training. We additionally tested the performance of HoVer-Net+ for layer segmentation following further training on nuclear segmentation/classification. For nuclear segmentation/classification we further compared HoVer-Net+ to HoVer-Net[9] and U-Net[8]. These models were trained based on their default parameters. For these comparisons, HoVer-Net and HoVer-Net+ were pretrained on PanNuke[10], and U-Net was pretrained on ImageNet. We further aimed to determine the generalizability of our HoVer-Net+ model to new datasets and different domain-invariance techniques.

*Results*

We aimed to test the generalisability of our proposed HoVer-Net+ model to new datasets in the presence of a domain shift (e.g. by scanner/site), when trained for layer segmentation alone. The data in C1 were collected on an Aperio scanner, whilst the data in C2 were predominantly collected on a Hamamatsu scanner. Thus, we expect a certain level of domain-shift between these cohorts. Our initial experiments therefore aimed to test our method, when trained/tested on combinations of the two Sheffield cohorts with annotations (C1 and C2). We also tested the effect of using different pretrained models (e.g. PanNuke[10] or ImageNet), and stain augmentation. The results for these experiments are displayed in Table S2. Training on both cohorts generally improved performance across both datasets, whereas stain augmentation was seen to not necessarily be beneficial. Overall, trial 5 gained the best overall F1-score, when pretrained on PanNuke and trained on both C1 and C2, without any stain augmentation. We suggest that when both datasets are present within both the train and test set, the model learns enough variation over the different scanners used, and thus stain augmentation is not necessarily useful.

Following optimisation, our HoVer-Net+ model (trained for layer segmentation) was compared with U-Net (no stain augmentation, trained on C1 and C2). The results for these comparative experiments can also be seen in Table S3. Here, multiple trials of HoVer-Net+ are displayed. First, HoVer-Net+$_{LS}$ describes the model trained for layer segmentation alone. Unsurprisingly, this method achieved the best F1-score of 0.819, whilst U-Net (ResNet-50 encoder) achieved the worst results for layer segmentation. We additionally make further comparisons, showing how the layer segmentation performance changes with the additional training of the HoVer-Net+ model with nuclear information. HoVer-Net+$_{LS,All}$ describes the model in its second stage of training. Following on from the first stage, where the model was trained for layer segmentation alone, the second stage describes the model that is then trained to perform simultaneous nuclear instance segmentation/classification and layer segmentation. Intuitively, there is a slight drop in the model performance, with incorporation of nuclear information, as there are less patches with both nuclear and layer information, and thus the model will be slightly overfitted to this smaller sample size. However, we combated this by introducing a third stage of training, where we froze the encoder, and the nuclear instance segmentation/classification decoder branches, and further tuned the LS decoder branch on the entire cohort of layer patches (HoVer-Net+$_{LS,All,LS}$). This acted to improve layer segmentation only slightly when compared to HoVer-Net+$_{LS,All}$.

The HoVer-Net+ results for nuclear instance segmentation and classification are included in Table S4; where we have additionally compared the performance against HoVer-Net[9] and U-Net[8]. U-Net has expectedly gained the worse performance, owing to its inability to differentiate touching nuclei well. This is best illustrated in Figure S1, where touching basal layer epithelial nuclei are poorly discriminated. Within this table, HoVer-Net+$_{LS,All}$ was initially trained on all layer segmentation data, and then further trained on all nuclei/layer data. Following the training of HoVer-Net+ on nuclei and layer information (e.g. HoVer-Net+$_{LS,All}$), unsurprisingly there is a decrease in performance for layer segmentation (see Table S4). The mean F1-score for HoVer-Net+ decreases from 0.819 to 0.812 owing to there being much fewer patches with both layer and nuclear annotations. We therefore opted to further finetune the HoVer-Net+ models (LS decoder only) on the entire cohort of layer patches (stage three of training). Following finetuning, the mean F1-score of HoVer-Net+ (HoVer-Net+$_{LS,All,LS}$ in Table S4) only increased marginally from 0.812 to 0.813. Since, this stage was done with the LS decoders alone, this should have little effect on the performance of HoVer-Net+ on nuclear segmentation/classification, only really effecting the nuclear classification performance. As can be seen from Table S4 this didn't affect the segmentation/detection quality of HoVer-Net+$_{LS,All,LS}$; however, did act to improve with the classification of nuclei. This demonstrates the benefits of using HoVer-Net+ in a multi-task learning framework. When we compare both of our HoVer-Net+ models to HoVer-Net we see an improved performance across both datasets, showing the superiority of the method over both cohorts C1 and C2. Thus, this model is used for inference on all WSIs in both the internal and external datasets seen in the next step of our analysis pipeline.

The output layer and nuclear segmentations of HoVer-Net+ on a ROI from the test set are displayed in Figure S1. The top row shows the GT layer segmentation on the left followed by HoVer-Net+ (HoVer-Net+$_{LS,All,LS}$) and U-Net. Here, we can see that HoVer-Net+ appears to have the best quality segmentation. The blue boxes show areas where HoVer-Net+ has segmented regions that the GT segmentation has misclassified. In general, U-Net's segmentations appear more spurious. The bottom row shows the nuclear segmentations for the GT, HoVer-Net+, HoVer-Net and U-Net (from left to right). Both HoVer-Net and HoVer-Net+ have incorrectly classified some epithelial nuclei as keratin nuclei, however, HoVer-Net appears to have misclassified many more of these cases, demonstrating the benefits of including the LS decoder branch.

*Discussion*

In the first stage of our experiment, we tested the performance and generalisability of our HoVer-Net+ method to new cohorts and domain invariance techniques. HoVer-Net+, when trained for layer segmentation, achieved the best results when pretrained on PanNuke over ImageNet. This is perhaps unsurprising, demonstrating the merits of pretraining on domain-specific images. We additionally tested how well HoVer-Net+ performed when trained/tested on different datasets. Overall, the mean F1-score across both C1 and C2 (used in training HoVer-Net+) was highest when trained over both datasets. Counterintuitively, we also found stain augmentation to have mixed effects on model performance, even when testing on new data. Since, we wished to generate a model that generalised as much as possible to new datasets we used the model trained on both datasets going forward, without any stain augmentation.

We performed multiple experiments testing the order of training HoVer-Net+ for each of the different tasks. We first trained the model on just layer segmentation data, as there were much more layer annotations than nuclear annotations. Next, we further trained the HoVer-Net+ model for nuclear segmentation/classification and layer segmentation (NP, HoVer, NC decoders pretrained on PanNuke, LS decoder trained in previous stage). Following this training (labelled HoVer-Net+$_{LS,All}$), there was an expected reduction in layer segmentation performance, owing to the encoder and LS decoder being further trained on a much smaller subset of data, and therefore being more susceptible to overfitting. We therefore circumvented this issue, by freezing the encoder and the nuclear decoder branches (NP, HoVer, NC), and finetuning the LS decoder branch on the entire layer segmentation data. This then resulted in an improved performance for layer segmentation (labelled HoVer-Net+$_{LS,All,LS}$) when compared to the HoVer-Net+$_{LS,All}$. Since the nuclear decoder branches were not altered here, this did not affect the nuclear segmentation performance of the model; but did improve the performance for nuclear sub-classification of epithelial nuclear types (e.g. as basal, epithelial, keratin), which is dependent on the output of the LS branch. This demonstrates the merits of our multi-task learning approach, where training over multiple tasks can improve the representation of the images learned by the encoder, to benefit both tasks.

Following the optimisation of HoVer-Net+ we compared its performance to U-Net for layer segmentation. Here, we found HoVer-Net+ to get the best performance, with U-Net being worse. We additionally compared HoVer-Net+ to HoVer-Net and U-Net for nuclear segmentation/classification, where we found HoVer-Net+ to gain the best results. Overall, these results express the benefits of our multi-task learning HoVer-Net+ method, which we expect to generalise well to new data.

**Table S2**

*Experiments for layer segmentation on both dataset 1 and 2, based on F1-scores.*

| Trial | Train Set | Pretrain. | Stain. Aug. | Test C1 Bkgd. | Other | Basal | Epith. | Keratin | Mean | Test C2 Bkgd. | Other | Basal | Epith. | Keratin | Mean | Overall Mean |
|---|---|---|---|---|---|---|---|---|---|---|---|---|---|---|---|---|
| 1 | C2 | PanNuke | N | 0.872 | 0.765 | 0.563 | 0.802 | 0.644 | 0.729 | 0.91 | 0.853 | 0.771 | 0.888 | 0.791 | 0.843 | 0.786 |
| 2 | C2 | PanNuke | Y | 0.822 | 0.713 | 0.557 | 0.739 | 0.547 | 0.675 | 0.909 | 0.828 | 0.714 | 0.861 | 0.774 | 0.817 | 0.746 |
| 3 | C1 | PanNuke | N | 0.842 | 0.807 | 0.717 | 0.819 | 0.660 | 0.769 | 0.844 | 0.813 | 0.720 | 0.851 | 0.780 | 0.802 | 0.785 |
| 4 | C1 | PanNuke | Y | 0.849 | 0.819 | 0.700 | 0.838 | 0.682 | 0.778 | 0.876 | 0.835 | 0.727 | 0.865 | 0.772 | 0.815 | 0.796 |
| **5** | **C1,C2** | **PanNuke** | **N** | **0.874** | **0.849** | **0.721** | **0.858** | **0.717** | **0.804** | **0.904** | **0.849** | **0.753** | **0.880** | **0.789** | **0.835** | **0.819** |
| 6 | C1,C2 | PanNuke | Y | 0.850 | 0.821 | 0.706 | 0.835 | 0.674 | 0.777 | 0.901 | 0.851 | 0.739 | 0.884 | 0.801 | 0.835 | 0.806 |
| 7 | C2 | ImageNet | N | 0.869 | 0.774 | 0.594 | 0.797 | 0.627 | 0.732 | 0.911 | 0.851 | 0.76 | 0.891 | 0.798 | 0.842 | 0.787 |
| 8 | C2 | ImageNet | Y | 0.843 | 0.742 | 0.544 | 0.781 | 0.603 | 0.703 | 0.91 | 0.842 | 0.695 | 0.868 | 0.776 | 0.818 | 0.760 |
| 9 | C1 | ImageNet | N | 0.878 | 0.849 | 0.715 | 0.852 | 0.698 | 0.798 | 0.864 | 0.473 | 0.533 | 0.446 | 0.251 | 0.513 | 0.656 |
| 10 | C1 | ImageNet | Y | 0.847 | 0.807 | 0.711 | 0.826 | 0.664 | 0.771 | 0.822 | 0.806 | 0.722 | 0.846 | 0.783 | 0.796 | 0.783 |
| 11 | C1,C2 | ImageNet | N | 0.864 | 0.808 | 0.701 | 0.815 | 0.654 | 0.768 | 0.905 | 0.777 | 0.714 | 0.809 | 0.578 | 0.756 | 0.762 |
| 12 | C1,C2 | ImageNet | Y | 0.817 | 0.755 | 0.672 | 0.780 | 0.577 | 0.720 | 0.888 | 0.834 | 0.713 | 0.870 | 0.784 | 0.818 | 0.769 |

**Table S3**

*Comparative experiments for layer segmentation on both dataset 1 and 2, based on F1-scores.*

| Model | Pretrain. | Test C1 Bkgd. | Other | Basal | Epith. | Keratin | Mean | Test C2 Bkgd. | Other | Basal | Epith. | Keratin | Mean | Overall Mean |
|---|---|---|---|---|---|---|---|---|---|---|---|---|---|---|
| U-Net | ImageNet | **0.880** | **0.849** | **0.739** | 0.853 | 0.708 | **0.806** | **0.905** | 0.777 | 0.714 | 0.809 | 0.578 | 0.756 | 0.781 |
| HoVer-Net+$_{LS}$ | PanNuke | 0.874 | **0.849** | 0.721 | **0.858** | **0.717** | 0.804 | 0.904 | **0.849** | 0.753 | 0.88 | 0.789 | 0.835 | **0.819** |
| HoVer-Net+$_{LS,All}$ | PanNuke | 0.862 | 0.837 | 0.722 | 0.846 | 0.707 | 0.795 | 0.902 | 0.834 | 0.759 | 0.872 | 0.78 | 0.829 | 0.812 |
| HoVer-Net+$_{LS,All,LS}$ | PanNuke | 0.853 | 0.827 | 0.716 | 0.851 | 0.700 | 0.789 | 0.898 | **0.849** | **0.764** | **0.885** | **0.792** | **0.837** | 0.813 |

*Note. HoVer-Net+$_{LS}$ was trained on layer data only. HoVer-Net+$_{LS,All}$ was trained on layer data only, then patches with both nuclei and layer data only. HoVer-Net+$_{LS,All,LS}$ was trained on layer data only, then patches with both nuclei and layer data only, finally, the LS decoder alone was finetuned on all layer data.*

**Table S4**

*Comparative experiments for nuclear instance segmentation and classification.*

| Model | Test C1 | | | | | | | | | Test C2 | | | | | | | | |
|---|---|---|---|---|---|---|---|---|---|---|---|---|---|---|---|---|---|---|
| | Dice | AJI | DQ | SQ | PQ | $F^d$ | $F_c^o$ | $F_c^b$ | $F_c^e$ | Dice | AJI | DQ | SQ | PQ | $F^d$ | $F_c^o$ | $F_c^b$ | $F_c^e$ |
| U-Net | 0.605 | 0.378 | 0.473 | 0.606 | 0.311 | 0.627 | 0.585 | 0.245 | 0.529 | 0.582 | 0.354 | 0.444 | 0.581 | 0.302 | 0.652 | 0.541 | 0.275 | 0.522 |
| HoVer-Net | 0.713 | 0.589 | 0.721 | **0.730** | **0.536** | **0.813** | 0.680 | 0.542 | 0.59 | 0.66 | 0.591 | 0.708 | 0.641 | 0.473 | 0.82 | 0.621 | 0.494 | 0.619 |
| HoVer-Net+$_{LS,All}$ | **0.716** | **0.598** | **0.724** | **0.730** | 0.533 | 0.810 | **0.719** | 0.599 | 0.633 | 0.665 | 0.635 | 0.753 | 0.641 | 0.478 | 0.833 | 0.715 | 0.59 | 0.646 |
| **HoVer-Net+$_{LS,All,LS}$** | **0.716** | **0.598** | **0.724** | **0.730** | 0.533 | 0.810 | 0.717 | **0.593** | **0.646** | **0.665** | **0.635** | **0.753** | **0.641** | **0.478** | **0.833** | **0.727** | **0.621** | **0.668** |

*Note. HoVer-Net+$_{LS,All}$ was trained on layer data only, then patches with both nuclei and layer data only. HoVer-Net+$_{LS,All,LS}$ was trained on layer data only, then patches with both nuclei and layer data only, finally, the LS decoder alone was finetuned on all layer data.*

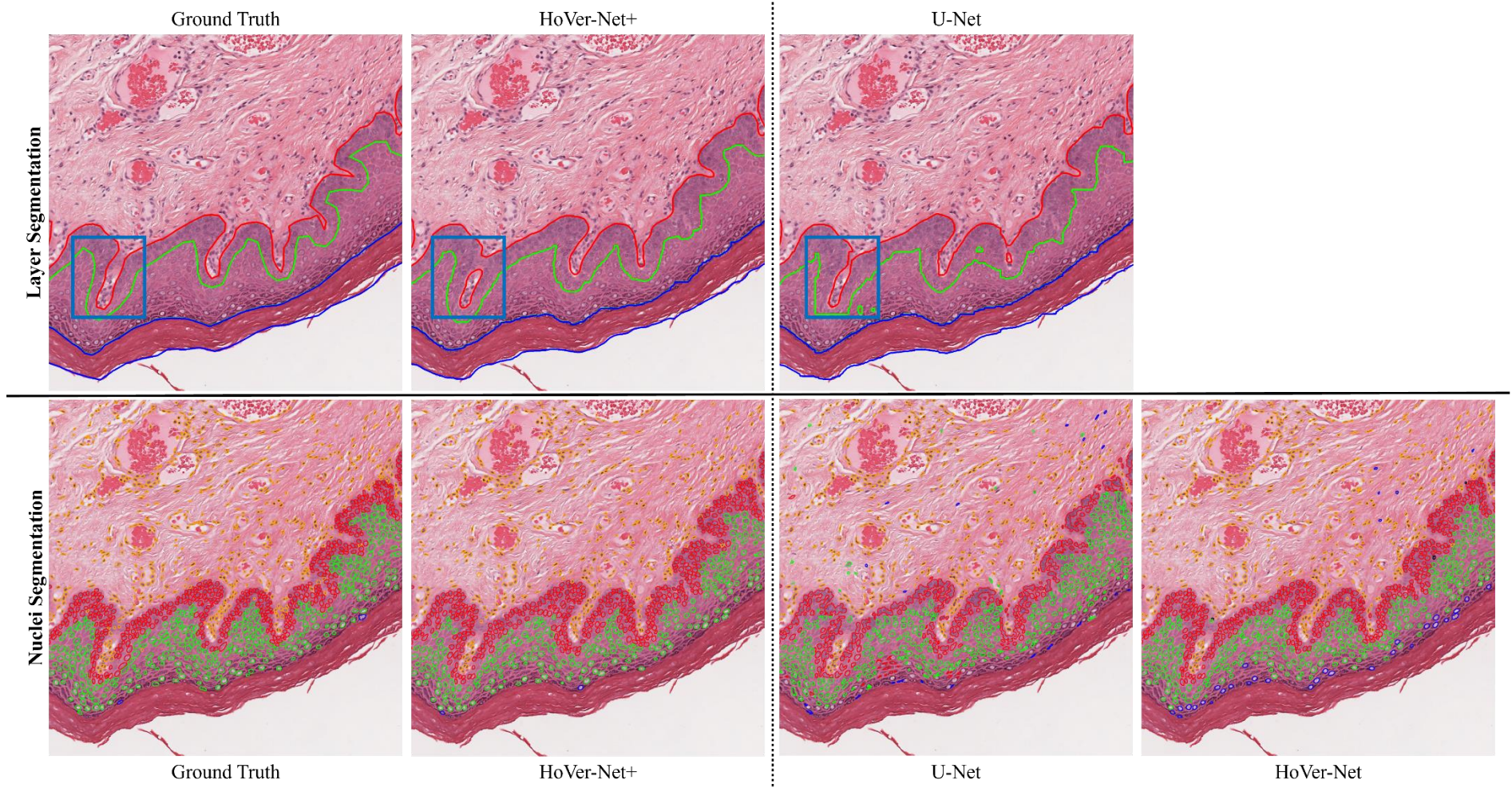

**Figure S1**
*Top row: A comparison of the GT layer segmentations vs. HoVer-Net+$_{LS,All,LS}$ and U-Net. In the images, the red line represents the basal layer, the green epithelium, and the blue keratin. For clarity, other tissue region segmentations have been excluded from these results. Bottom row: A comparison of the GT nuclear segmentations vs. HoVer-Net+$_{LS,All,LS}$, HoVer-Net and U-Net. In the images, the red line represents the basal nuclei, green epithelial nuclei, blue keratin nuclei, orange other nuclei and finally, black represents unlabelled nuclei. Note, the dividing line between the annotations on the left and right is to display that the HoVer-Net+ segmentations for nuclei and layers are from the same model. On the right, specifically with U-Net, these are from two separately trained models.*

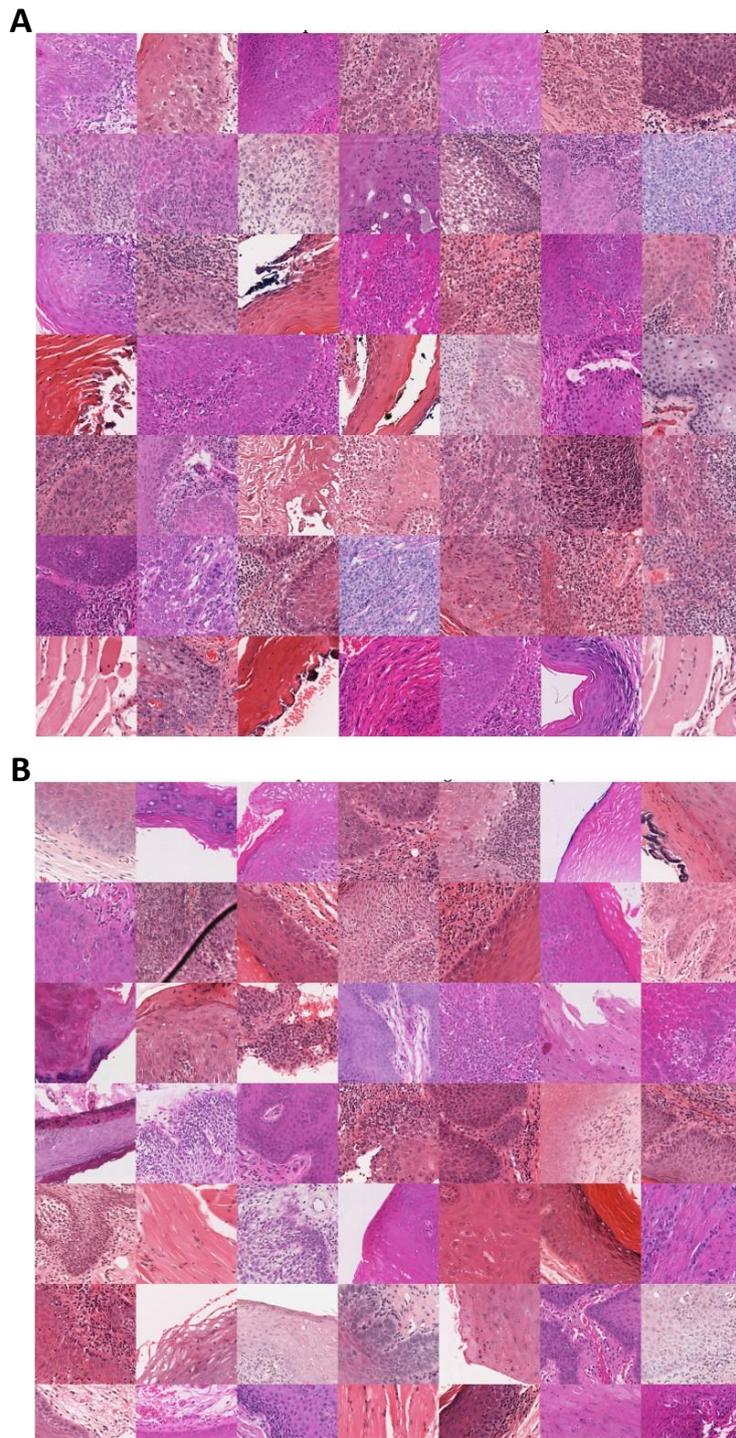

**Figure S2**
*Random patch montage of A) top predicted patches in true positives and B) bottom predicted patches in true negatives from our MLP.*

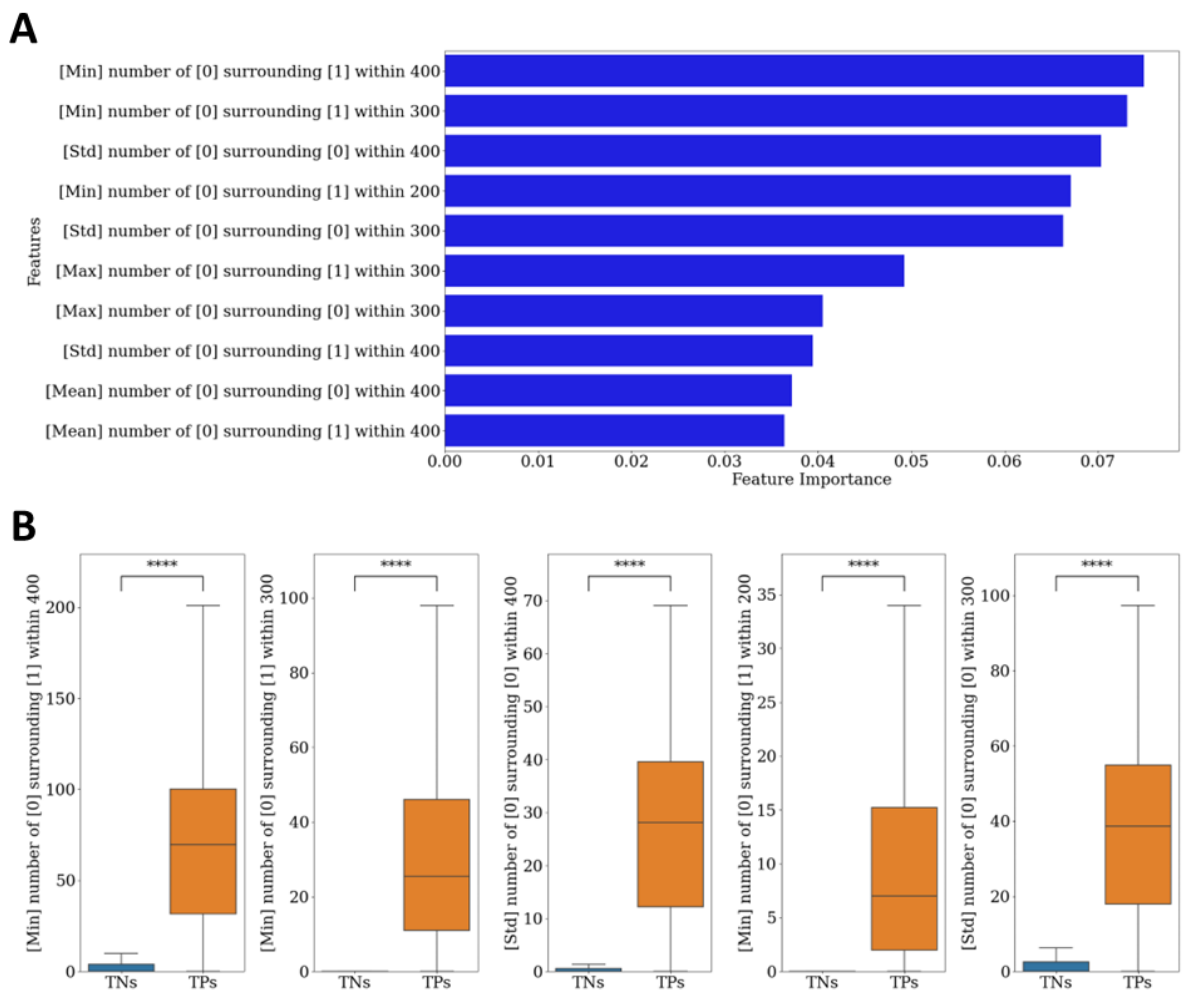

**Figure S3**
*A) The top ten most important morphological/spatial features found via a Random Forest, and B) bosplots showing the distribution of the top five of these features. [0] are "other" nuclei, whilst [1] are epithelial nuclei. Distance is measure in microns.*

**Slide-level Transformation Prediction**

*Nuclear Features*

We generated tile-level features for use in our MLP model for transformation prediction. For each tile, we calculated 104 morphological and 64 spatial features. The morphological features were obtained from 13 shape features for each nucleus in a tile (eccentricity, convex area, contour area, extent, perimeter, solidity, orientation, radius, major/minor axis, equivalent diameter, bounding box area/aspect ratio) with four tile-level statistics (mean, minimum, maximum, variance) per nuclear type (epithelial and other). This resulted in 104 morphological features per tile. We computed the number of different nuclear types within a small radius of a nuclear instance, resulting in four counts per tile (number of epithelial nuclei around another nucleus, number of epithelial nuclei around epithelial nuclei, number of other nuclei around epithelial nuclei, and finally the number of other nuclei around other nuclei) over four varying radii (100, 200, 300 and 400 pixel radii). Finally, we took tile-level summary statistics (mean, minimum, maximum, standard deviation) across these 16 features, resulting in 64 spatial features per tile.

*MLP Coldspot/hotspot Patches*

For slide-level prediction, we used a multi-layer perceptron (MLP), trained with the iterative draw-and-rank method introduced by Bilal *et al.* (2021)[36] with our tile-level morphological/spatial features. We display a randomly generated selection of the top five most predictive patches for TP cases and bottom five patches from TN cases on internal testing in Figure S2. Visually, it is clear that there appear to be a larger number of immune cells both within the epithelium and the connective tissue within the TP cases.

*Hotspot-Coldspot Feature Importance*

We additionally aimed to determine which of the 168 nuclear features used to train our MLP were most important for making the final prediction (on internal testing). Thus, we trained a Random Forest classifier based on the 168 nuclear features in the top five/bottom five predicted patches by our MLP model. We then ranked the feature importance (mean decrease in impurity, MDI), selected the top ten features and performed two-tailed t-tests (with FDR correction), to determine statistical significance.

We present the top ten features found from this analysis (using MDI importance) in Figure S3A. Spatial features were generally estimated to be most important and made up the top ten features. We further display boxplots showing the distribution of the top five of these features in in Figure S3B and provide results from a two-tail t-test (following FDR correction). The minimum number of "other" nuclei surrounding an epithelial nucleus within 400 microns had the highest importance, although all of the top ten features had large effect values ($d > 2.00$) and were shown to be significant ($p < 0.0001$). These top features again suggest the importance of lymphocytes (e.g. "other" nuclei) and their proximity to epithelial nuclei (e.g. epithelial-infiltrating/peri-epithelial lymphocytes).

*Hotspot-Coldspot Feature Exploration*

We did not incorporate any mitotic features into our models, however, we additionally aimed to determine whether our top/bottom predicted patches could further inform us of their importance. We therefore, ran the MIDOG 2022 winning code (available online: [adamshephard/TIA-mitosis (github.com)](adamshephard/TIA-mitosis)) on the top/bottom predicted patches from our analyses. Since mitotic figures are generally sparse in dysplasia, we generated a mitotic count per slide over all of the five top/bottom predicted patches by the model (in TP vs TN cases, respectively). Only slides that included one or more mitotic figures were included in this analysis. A two-tailed t-test was then used to determine statistical significance.

The MIDOG 22 winning model appeared to work well on our dataset (by visual inspection). However, we found no difference between the number of mitotic figures discovered in TPs vs TNs ($d = 0.06$, $p = 0.82$). This was perhaps an unsurprising finding, as mitotic figures were not included as a feature in our MLP models.

*MLP External Testing*

We additionally aimed to test the performance of our MLP model, on external validation, without the exclusion of cases that HoVer-Net+'s nuclear/region segmentation results were deemed insufficient quality. The results of this analyses are shown in Table S5. We see a clear drop in performance compared to with the QC of nearly 10%. This shows the importance of having a robust segmentation pipeline for this analysis. With training on more diverse data, we expect HoVer-Net+ to generalise better to unseen datasets.

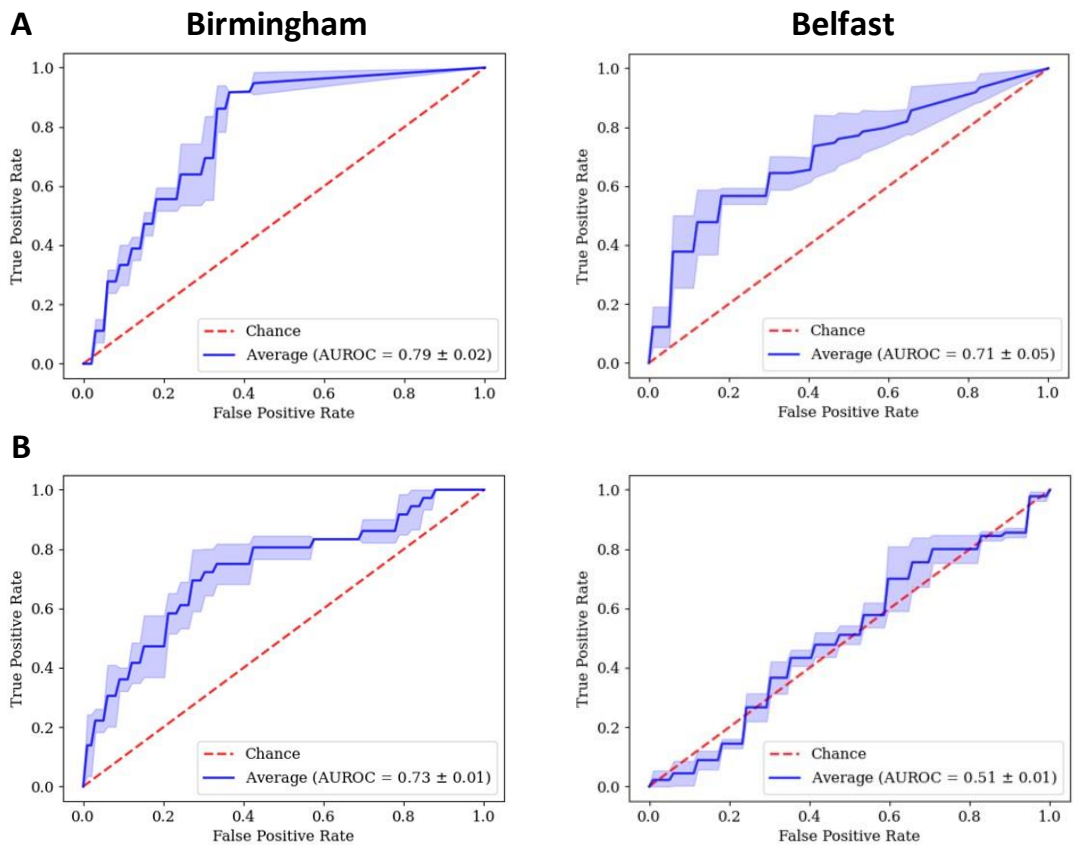

**Figure S4**
*AUROC curves for the A) MLP vs B) IDaRS-SA-DA on external testing.*

*Comparison to IDaRS for Transformation Prediction*

We additionally compared our MLP method for transformation prediction to another state-of-the-art method. Thus, we compared it to IDaRS[11], trained with deep features, in its original published form. We suggest that one of the main advantages of our MLP trained with morphological/spatial features, is that these features are in theory "domain-agnostic". Thus, providing the deep learning models used to generate the nuclear segmentations generalise, then these features should generalise well to new, unseen datasets. In contrast, we suggest that deep models such as IDaRS may not generalise well to unseen data. Thus, we have additionally employed domain adaption techniques to make the IDaRS model robust to new data, in order to make a fair-comparison of the IDaRS method to our MLP. Thus, we additionally include the IDaRS model, trained using a) Macenko stain augmentation, b) domain adversarial training, and c) a combination of these two techniques. Domain adversarial training was introduced by Ganin *et al.* (2016)[12], and aims to force the encoder part of the CNN to learn domain-agnostic features. Within this work, an additional classifier head is added to the IDaRS model that aims to predict the domain from which the provided features are from (in this case, the scanner used).

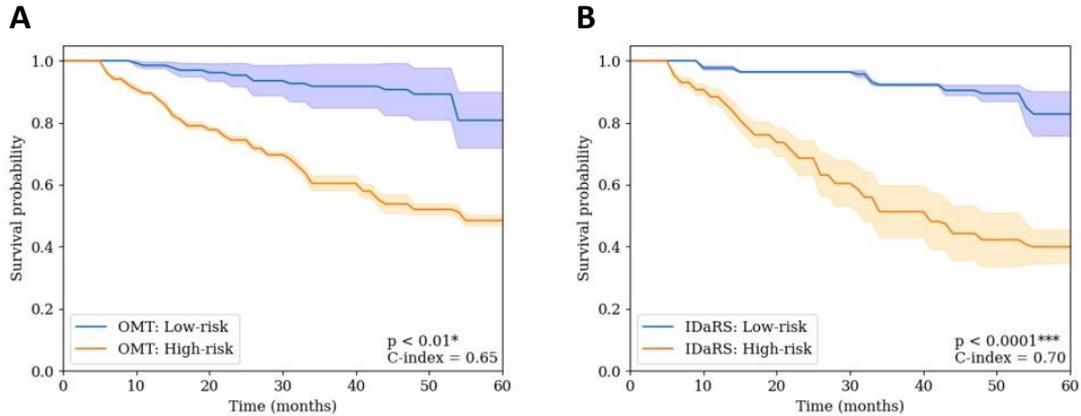

**Figure S5**
*Kaplan-Meier transformation-free survival curves for A) the MLP produced OMT-grade, and B) IDaRS on the external Birmingham-Belfast dataset.*

The addition of a gradient reversal layer aims to maximise the error in this classification task, thus incentivising the classification model to learn domain-agnostic features.

In all models presented in this work, the patches used for feature generation are based on HoVer-Net+'s segmentation of the epithelium. All models were additionally trained with a symmetric cross-entropy loss function, and the Adam optimiser. We chose the parameters $k = 5$ top predictive patches and $r = 45$ random patches in the iterative draw-and-rank method, with a batch size of 256. All models were trained until convergence with a minimum of 30 and a maximum of 100 epochs.

In these experiments (see Table S6), we have compared the performance of our MLP trained with patch-level morphological/spatial features to the IDaRS model (based on deep features from a ResNet-34 model). One of the main advantages of our MLP trained with morphological features, is that these features are in theory "domain-agnostic", thus, we additionally compare our model to IDaRS trained with a) stain augmentation (IDaRS-SA), b) domain adversarial training (IDaRS-DA), and c) IDaRS trained with both stain augmentation and domain adversarial training (IDaRS-SA-DA).

Overall, the results tell an interesting story, with relation to domain adaption. The raw IDaRS model that obtained the highest results on internal cross-validation (AUROC = 0.78), has not generalised well to the new unseen data (AUROC = 0.64). The various domain adaption techniques, such as domain adversarial training and stain augmentation, did not help to improve model performance on internal testing; but, helped improve the generalisability of the IDaRS model on external testing. However, ultimately, the model that was trained with domain-agnostic features (e.g. our MLP) achieved the highest results on external testing (AUROC = 0.74). We additionally show the AUROC curves for these models in Figure S4. These plots clearly demonstrate the superiority of the MLP to IDaRS-SA-DA, especially in relation to the Belfast data. The Kaplan-Meier transformation-free survival curves for these two models are also shown in Figure S5 for the Birmingham-Belfast cohort. Interestingly we see an improvement of IDaRS over the MLP, particularly in relation to smaller confidence intervals.

**Table S5**
*Slide-level mean (standard deviation) results for transformation prediction when trained on the Sheffield cohorts and tested on the external Birmingham-Belfast data, without any QC of HoVer-Net+ output.*

|  |  | Birmingham | | Belfast | | Combined | |
| --- | --- | --- | --- | --- | --- | --- | --- |
| Model | Features | F1-score | AUROC | F1-score | AUROC | F1-score | AUROC |
| **MLP** | **Morph./Spatial** | 0.59 (0.05) | 0.66 (0.01) | 0.64 (0.10) | 0.65 (0.05) | **0.59 (0.03)** | **0.64 (0.01)** |
| Binary Grade | Low- vs High-Risk | | | 0.74 (0.00) | 0.52 (0.00) | | |
| WHO Grade | Mild/Mod. vs Severe | 0.36 (0.00) | 0.60 (0.00) | 0.35 (0.00) | 0.40 (0.00) | 0.34 (0.00) | 0.53 (0.00) |
| WHO Grade | Mild vs Mod./Severe | 0.51 (0.00) | 0.73 (0.00) | 0.73 (0.00) | 0.49 (0.00) | 0.60 (0.00) | 0.64 (0.00) |

**Table S6**
*Slide-level mean (standard deviation) results for transformation prediction.*

|  |  | Internal Testing | | External Testing | | | | | |
|  |  | Sheffield | | Birmingham | | Belfast | | Combined | |
| --- | --- | --- | --- | --- | --- | --- | --- | --- | --- |
| Model | Features | F1-score | AUROC | F1-score | AUROC | F1-score | AUROC | F1-score | AUROC |
| IDaRS | ResNet-34 | 0.59 (0.11) | 0.78 (0.11) | 0.50 (0.09) | 0.67 (0.04) | 0.72 (0.06) | 0.57 (0.04) | 0.63 (0.05) | 0.64 (0.02) |
| **IDaRS-SA** | **ResNet-34** | **0.59 (0.07)** | **0.80 (0.07)** | 0.48 (0.10) | 0.63 (0.08) | 0.62 (0.05) | 0.56 (0.04) | 0.64 (0.03) | 0.69 (0.04) |
| IDaRS-DA | ResNet-34 | 0.55 (0.10) | 0.73 (0.11) | 0.56 (0.06) | 0.74 (0.04) | 0.72 (0.06) | 0.50 (0.02) | 0.70 (0.02) | 0.69 (0.02) |
| IDaRS-SA-DA | ResNet-34 | 0.53 (0.08) | 0.72 (0.11) | 0.57 (0.06) | 0.71 (0.02) | 0.64 (0.12) | 0.51 (0.02) | 0.63 (0.06) | 0.69 (0.02) |
| **MLP** | **Morph./Spatial** | 0.57 (0.08) | 0.77 (0.08) | **0.64 (0.01)** | **0.78 (0.02)** | 0.73 (0.09) | 0.71 (0.06) | **0.70 (0.05)** | **0.74 (0.08)** |
| Binary Grade | Low- vs High-Risk | 0.51 (0.08) | 0.71 (0.06) | - | - | **0.74 (0.00)** | 0.52 (0.00) | - | - |
| WHO Grade | Mild/Mod. vs Severe | 0.34 (0.16) | 0.58 (0.11) | 0.44 (0.00) | 0.64 (0.00) | 0.35 (0.00) | 0.40 (0.00) | 0.36 (0.00) | 0.54 (0.00) |
| WHO Grade | Mild vs Mod./Severe | 0.46 (0.08) | 0.68 (0.05) | 0.61 (0.00) | 0.75 (0.00) | 0.73 (0.00) | 0.49 (0.00) | 0.64 (0.00) | 0.64 (0.00) |

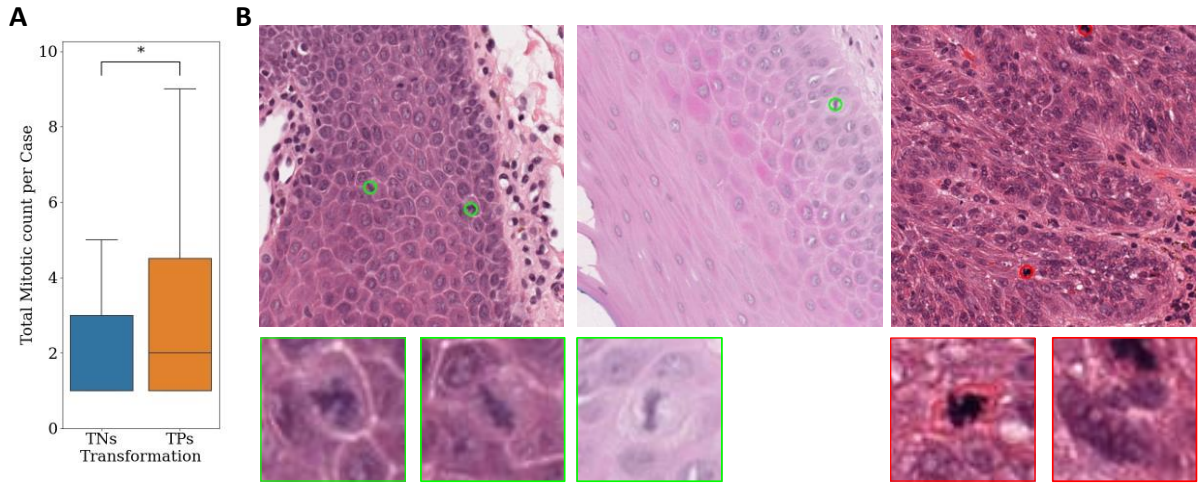

**Figure S6**
*A) Mitotic figure counts (MIDOG 2022 winning algorithm), in the top five predicted patches from true positive cases against the bottom five predicted patches from true negative cases. B) Detections with algorithm correctly detected mitoses (green) and incorrectly missed mitoses (red).*

### IDaRS Feature Discovery

IDaRS generated predictions based on deep features learned through training (here, ResNet-34). This model differs from the MLP pipeline used primarily in this study, in that the features that are used in the MLP are deliberately generated, with the purpose of being interpretable and important. However, the IDaRS framework does generate good predictions for both internal validation, and external validation (with domain adaption techniques). Thus, we also explored the hot/coldspots found by the IDaRS model (on internal testing, IDaRS-DA-SA) to determine what were the important features used for prediction. Interestingly, we found a significantly higher number of mitotic figures in the TP patches than the TN patches ($d = 0.63$, $p = 0.01$; see Figure S6). This suggests that the IDaRS model has independently learned to account for mitosis as a deep feature.

### Discussion

In this work, we aimed to generate a model that would better predict transformation, using interpretable features. Ultimately, these morphological/spatial features were based on shape/size differences (e.g. the differences in the mean radius of nuclei in a patch) in different nuclear types (e.g. epithelial vs other nuclei). Overall, we found our MLP model to gain the best results for predicting transformation on external testing. Interestingly, despite the success of IDaRS with deep features in the past, we have found that by instead using morphological/spatial features one can achieve better performance. We also argue that the incorporation of such features in favour of deep features, makes our models more interpretable. IDaRS, which initially gained the highest AUROC values on internal testing, failed to generalise to the new, unseen data without any form of domain alignment/augmentation techniques. As was expected, the addition of stain augmentation and domain adversarial training, improved the performance of IDaRS on new data (from AUROC = 0.64 to AUROC = 0.69). However, ultimately our MLP model, trained on interpretable morphological/spatial features gained the highest performance (AUROC = 0.74). These features are essentially domain-agnostic (when segmented accurately), and we suggest that this is the reason for their superior performance.

Despite this, we still explored the hotspot/coldspot regions produced by IDaRS, and found that mitoses potentially contributed to the model performance. Future work should look at incorporating tile-level (or even slide-level) mitotic scoring into the MLP model, to further improve the performance.

**Supplementary Material References**

1. Alemi Koohbanani, N., Jahanifar, M., Zamani Tajadin, N. & Rajpoot, N. NuClick: A deep learning framework for interactive segmentation of microscopic images. *Med. Image Anal.* **65**, (2020).

2. Jahanifar, M., Koohbanani, N. A. & Rajpoot, N. NuClick: From Clicks in the Nuclei to Nuclear Boundaries. *arXiv* (2019).

3. Graham, S. *et al.* One model is all you need: Multi-task learning enables simultaneous histology image segmentation and classification. *Med. Image Anal.* **83**, (2023).

4. Pocock, J. *et al.* TIAToolbox as an end-to-end library for advanced tissue image analytics. *Commun. Med.* **2**, 120 (2022).

5. Aubreville, M. *et al.* Mitosis domain generalization in histopathology images — The MIDOG challenge. *Medical Image Analysis* vol. 84 102699 (2023).

6. Jahanifar, M. *et al.* Stain-Robust Mitotic Figure Detection for the Mitosis Domain Generalization Challenge. *arXiv* 3–5 (2021).

7. Jahanifar, M., Shephard, A., Zamanitajeddin, N., Raza, S. E. A. & Rajpoot, N. Stain-Robust Mitotic Figure Detection for MIDOG 2022 Challenge. 1–3 (2022).

8. Ronneberger, O., Fischer, P. & Brox, T. U-net: Convolutional networks for biomedical image segmentation. *Lect. Notes Comput. Sci. (including Subser. Lect. Notes Artif. Intell. Lect. Notes Bioinformatics)* **9351**, 234–241 (2015).

9. Graham, S. *et al.* Hover-Net: Simultaneous segmentation and classification of nuclei in multi-tissue histology images. *Med. Image Anal.* **58**, 101563 (2019).

10. Gamper, J. *et al.* PanNuke Dataset Extension, Insights and Baselines. 1–12 (2020).

11. Bilal, M. *et al.* Development and validation of a weakly supervised deep learning framework to predict the status of molecular pathways and key mutations in colorectal cancer from routine histology images: a retrospective study. *Lancet Digit. Heal.* **3**, e763–e772 (2021).

12. Ganin, Y. *et al.* Domain-adversarial training of neural networks. in *Advances in Computer Vision and Pattern Recognition* vol. 17 189–209 (2017).